\documentclass[10pt,letterpaper]{article}

\usepackage[top=0.85in,left=2.75in,footskip=0.75in]{geometry}
\usepackage{amsmath,amssymb}
\usepackage{changepage}
\usepackage{textcomp}
\usepackage{cite}
\usepackage{nameref,hyperref}
\usepackage[nopatch=eqnum,expansion=false]{microtype}  
\DisableLigatures[f]{encoding = *, family = * }
\usepackage[table]{xcolor}
\usepackage{array}
\usepackage{caption}
\usepackage{lastpage,fancyhdr,graphicx}
\usepackage{epstopdf}

\usepackage{amsfonts}
\usepackage{placeins}               
\usepackage{adjustbox}              
\usepackage{longtable}
\usepackage{colortbl}
\usepackage{multirow}
\usepackage{booktabs}
\usepackage[linesnumbered,ruled,vlined]{algorithm2e}
\usepackage[utf8]{inputenc}
\usepackage[T1]{fontenc}

\makeatletter
\renewcommand{\@biblabel}[1]{\quad#1.}
\makeatother

\SetCommentSty{mycommfont}
\SetKwInput{KwInput}{Input}
\SetKwInput{KwOutput}{Output}

\date{}

\begin{document}
\vspace*{0.2in}

\begin{flushleft}
{\Large
\textbf\newline{Scientific Data Analysis for Class-Informatics in Computational Taxonomy}
}
\newline
\\
Michelon, Thomas B.\textsuperscript{1},
Fushing Hsieh\textsuperscript{2,3,*}
\\
\bigskip
\textbf{1} Department of Plant Science, Federal University of Paran\'{a}, Curitiba, PR, Brazil.
\\
\textbf{2} Department of Statistics, University of California at Davis, CA 95616, USA.
\\
\textbf{3} Department of Statistics, University of California at Davis, CA 95616, USA.
\\
\bigskip
* Corresponding author\\
E-mail: fhsieh@ucdavis.edu
\end{flushleft}

\section*{Abstract}
In this A.I. era, Computational Taxonomy is proposed to study complex systems by analyzing their databases under taxonomic hierarchies abiding the Principle of Science by providing ``good explanations''. Comparisons among branches or classes are carried out by Scientific Data Analysis (SDA) paradigm that explores all potential associative patterns, including interacting effects of all high orders, and evaluate finite sample precisions for all information pieces individually by effectively making use of all variables' categorical nature. Under each comparison, all confirmed information pieces are collected and displayed along row-axis of a heatmap with all involved study-subjects on the column-axis. Each comparison's heatmap individually characterizes participating classes and study-subjects and simultaneously provides a scientific basis for outlier detection upon all non-participants. All these heatmaps then collectively constitutes so-called Class-informatics that offers good explanations based on characteristic of all classes and study-subjects. Computational Taxonomy's Class-informatics indeed resolves multiple fundamental issues: Tukey's more than 60 years outlier detection problem, issue of self-correction annotation, and a crucial check on assumption of information-content equality between testing and training data sets in Machine Learning. A showcase of Computational Taxonomy is exclusively illustrated on Iris data.

\section{Introduction}
\subsection{Propos and Prelude}
``The quest for good explanations is, I believe, the basic regulating principle not only of science, but of the Enlightenment generally.''

Quantum Computing physicist David Deutsch wrote this quote in the first chapter of his book: ``The beginning of Infinity: Explanations That Transform the World'' \cite{deutsch11}. Further, he emphasized that only such hard-to-varying ``good explanations'' can be translated into real progresses in society. An excellent example shedding lights on this principle is the greatest data analysis performed by Johannes Kepler (1571-1630).

In history of science, Kepler first worked with Tycho Brahe to make and record observations on planetary positions over time. Kepler even constructed his own table of logarithms to expedite his astronomical calculations with Napier’s invention of logarithms. After years of laborious hand calculations for compression of the data, he arrived at his magnificent his three laws of planetary motion (elliptical orbits, equal areas in equal times, orbital period relation) \cite{russell}.

As Johannes Kepler described how planets move with his three laws, Isaac Newton explained why, showing these patterns arise from his Law of Universal Gravitation \cite{markowsky}. In other words, Isaac Newton transformed Kepler's observational descriptions into a physical theory by building directly upon Kepler's empirical data, demonstrating that gravity provides the centripetal force for orbits and refining Kepler's Third Law by incorporating planetary masses.

In view of this historical example, nowadays the phrases ``data analysis'' and ``Data Science'' are apparently used rather carelessly to advance various agenda in industry as well as on campus. Almost 15 years ago, the first chapter of David Deutsch's book contained one message that gives an accurate description about how data analysis is perceived and used in current science community.  On page 15, ``..... In some fields (such as statistical analysis) the very word 'explanation' has come to mean prediction, so that a mathematical formula is said to 'explain' a set of experimental data. By 'reality' is meant merely the 'observed data' that the formula is supposed to approximate.'' It is reasonable to call this wide-spreading practice in sciences and industries:``the modeling-prediction doctrine.''

David Deutsch pushed the following messages to reject this modeling-prediction authority. On page 14, ``....that prediction is not, and cannot not be, the purpose of science.''. And on page 18, ``... Again, the very term 'data' ('givens') is misleading. Amending the 'data', or rejecting some as erroneous, is a frequent concomitant of scientific discovery, and the crucial 'data' cannot even be obtained until theory tells us what to look for and how and what.'' Further, to encourage scientists to wake up from this unscientific doctrine as another form of authority, he cited a quote from physicist Richard Feynman: ``Science is what we have learned about how to keep from fooling ourselves''.

Contrasting to these messages, the current state of science says quite the opposite. That is, the modeling-prediction doctrine might have driven many branches of sciences into a metastasis state. When facing this metastasis in sciences, how can data analysis in sciences follow the principle of science?

Data analyses for sciences indeed require a great leap of imagination on top of computing power. Since, imaginary capacities are critically needed for intricate data compression, especially in feature selections, in order to reveal explanatory nature. If such capacities indeed depend on domain knowledge, then shouldn't data analysts at least try their best to pave ways for domain scientists to exercise their imaginations? This partnership between data analysts and domain scientists might be remedial for the metastasis in science. To build such partnership, the best approach is to uphold this quest for good explanations when analyzing data. In this paper, Computational Taxonomy (CT) via Scientific Data Analysis (SDA) is proposed as an universal theme to up-hold this quest for better understanding complex systems.

\subsection{Computational Taxonomy and Scientific Data Analysis}
Nowadays, the classification is ubiquitous in the fields of Statistics, Machine Learning and A.I. Its task is narrowly defined as computationally ``predicting'' test-subject's class via classifiers such as variants of Random Forest and Boosting and Deep Learning, among many others. Among all involved classes nothing is known about: ``Which classes are closer to which, but far away from which''. The need of building a taxonomic hierarchy upon all classes is hardly a concern at all. Hence, classification results are reported in a black-box fashion without explanations. As David Deutsch even put it: `` A prediction without good explanations is indeed a prophesy.'' \cite{deutsch97}. Such a predictive operation is basically unscientific.

Further, the validity of such predictive classification is seemingly justified via classification errors. Its unscientific fabric is exposed as that a predictive result is given without checking the prerequisite: ``outlier status''. By ignoring this prerequisite, any accuracy loses its true meaning. Furthermore, all errors are equal. Again this false phenomenon of homogeneity is caused by ignoring the existential taxonomic hierarchy representing system's intrinsic heterogeneity. Heterogeneity is known as one natural fabric of real-world complex system \cite{anderson}.

Historically and functionally speaking, taxonomy and classification are two sides of one science: 1) building an underlying hierarchy of classes, so-called taxonomy; 2) allocating study subjects to the classes, so-called classification. In taxonomy, taxonomic units known as "taxa" are typically organized to receive a coordinate within the hierarchical construct that would precisely lay out varying Signal-to-Noise(S-N)ratios for performing classification tasks to retrieve such coordinates.

In this paper, we adopt a term: Class-informatics, to stand for all organized data-driven information contents under a taxonomy representing a targeted complex system. Originally, the taxonomic and classification tasks indeed are performed by one scientist called Taxonomist. As such, when Class-informatics is equipped with various Application Programming Interfaces (API) and Large language Model (LLM), it would act like a virtual taxonomist, who can answer all system related questions.

In contrast, where are taxonomists in biological sciences and beyond? Unfortunately, the speciality is nearly extinct from academic campus. It is time for data analyst step up. That is, the modern taxonomist must work on complex systems with living organisms and non-living entities. As such, hopefully, data-driven scientific understanding and explanations can be translated into technological progresses within our society in future. This modern Taxonomy is indeed even more important than ever in this Big Data and A.I. eras. Since people on the receiving end can only ``trust'' information with good explanations derived from scientifically conducted data analysis.

Therefore, it is worth iterating the essential merits of Class-informatics from the following four perspectives. First, experts and data curators of the targeted complex system are unlikely equipped with the complete knowledge regarding the entire system of multiscale nature, but a computer can. Since a system's representative taxonomic hierarchy would facilitate multiscale understanding onto systemic heterogeneity that surely embedded within any targeted complex system \cite{anderson}.

Secondly, upon details of varying scales, there are rooms for corrections on annotated-IDs here and there. Class-informatics built by Scientific Data Analysis (SDA) based Computational Taxonomy would offer a self-correcting mechanism for correcting annotated class-IDs. Thirdly, Class-informatics offers an explicit resolution to the hugely important, but still not-yet-resolved, fundamental issue in Machine Learning and A.I. literature: How to check whether the $20\%$ testing data set retains the same information content as the $80\%$ training data set? Without confirming the validity of this assumption, there is no way to claim that machine learning methodologies are scientific.

Lastly, Class-Informatics resolves the aforementioned fundamental issue: How to see and perform outlier detection? Though outlier detection is the prerequisite of any prediction, this issue is still standing after more than 60 years when it was first identified by John Tukey in his 1962 paper with title: " The Future of Data Analysis"\cite{tukey}. He also emphasized repeatedly in the paper that Data Analysis must be a scientific discipline for all sciences.

In summary, this paper follows John Tukey's this direction by proposing SDA based Computational Taxonomy and creating Class-informatics to expand trustworthy scientific results and authentic characteristics. They together fulfill the quest for good explanations pertaining to the complex system under study. Hopefully,  data analysts could help and motivate domain scientists to conjecture new knowledge and intelligence. As such technologies are eventually advanced to make critical progresses within our human society.

\section{Fundamental ingredients of Taxonomic Hierarchy and Class-informatics.}
\paragraph{Regarding a spectrum of response (Re) variables:} How should data analysts proceed when exploring a classification data set without an established Taxonomic Hierarchy? Since this hierarchical structure or geometry is supposed to be built upon a collection of annotated class-IDs pertaining to a complex system of interest. It would need, multiple categorical response variables, not just one, to be designated, defined and explored. This proceeding protocol is due to the fact that S-N ratios vary in comparisons involving different subsets of class-IDs. Ideally, such data analyses upon different Re-Co dynamics could collectively reveal the taxonomy of complex system as one whole.

The needed spectrum of response variables must include all possible pairwise comparisons of two class-IDs and the one of full collection of class-IDs. Though the response variable consisting the whole set of class-IDs likely sheds some lights on the taxonomic hierarchy, those computed patterns are subject to relatively lower S-N ratios. Hence, it would need all possible pairwise comparisons, making sure to include those ones with high S-N ratios, to confirm the most basic tree-hierarchy for the taxonomic task. Analysis across this spectrum of response variables would also bring out varying importance of different covariate (Co) features and feature-sets in different comparisons.

Without loss of generality, here the targeted taxonomic hierarchy would takes the binary tree-geometric format for expositional simplicity. So that all response(Re) variables are branch-based  categorical comparisons. With all covariate(Co) features, which are properly categorized and fused if necessary to go into SDA computations, one response variable makes up one Re-Co dynamics. We expect that this collection of Re-Co dynamics jointly represents all critical facets of complex system dynamics.

\paragraph{Regarding Class-informatics for a spectrum of Re-Co dynamics:} Given any designated categorical response variable, the corresponding Re-Co dynamics is supposed to be fully characterized by possible associative relational information from Co-to-Re. Hence, each piece of associative information provided by categories of a covariate feature or feature-set will bear with classifying functions as a fabric of taxonomy. Collectively, such associative information pieces will depict key aspects of characteristics of each involved Class-IDs. Further, this collective of associative information pieces surly would embrace all involved study-subjects' individual characteristics from the perspective of this Re-Co dynamics. Such idiosyncratic characterization of any involved study-subject is called individual character-landscape. As such similarity and dissimilarity among participating study-subjects can be measured and their topological neighborhood system can be built. This topology is essential for Classification task.

Furthermore, the most surprising fact is that such a collective of associative information pieces also provide a basis for characterizing non-participating Classes' study-subjects through ``outlier detection'' with respect to participating study-subjects' topological neighborhood system. That is, each non-participating study-subject belonging to the complementary sub-collection of Class-IDs will have its outlier status established. Thus, such a Re-Co dynamics is characterized by the synthesis of such collectives of associative information pieces and their corresponding collectives of outlier statues. Any classification decision-making within such a Re-Co dynamics is explainable, even visible.

Finally, a spectrum of Re-Co dynamics would shed lights on which classes is closer to which, but far away from which. As such the hierarchy upon the collection of Class-IDs is exposed and then established. This constructed hierarchy is the task of taxonomy. In turn, this hierarchy is the road-map for carrying out the task of classification. This is how Computational Taxonomy is performed based on a classification data set. As such the systematic collective of multiscale: study-subjects' and class-specific, characteristics information is generically called ``Class-informatics'' of this data set. For future scientific literatures, such taxonomy-oriented Class-informatics would be truly critical.

In view of scientific goal and societal implications, Class-Informatics is seen by embracing individual characteristics of all involving classes with holistic intrinsics of underlying dynamics of the targeted complex system as one whole. Ideally Class-informatics can be taken as computationally restored knowledge and intelligence regrading the complex system and beyond. The word ``beyond'' is specifically referred to high order interacting effects that most often point to under-explored sub-domains of the complex system. So, they are potential bases of ``conjectures'' for further and better understanding onto the system of interest.

\paragraph{Regarding SDA:} Within Computational Taxonomy, the engine for constructing Class-informatics is the Scientific Data Analysis(SDA). That is, in a completely data-driven fashion, SDA explores all possible associative relational pattern information pieces and then confirm reliable ones within each Re-Co dynamics identified within the construction of Taxonomic Hierarchy and all branch-vs-branch comparisons after the construction. Under a Re-Co dynamics, SDA performs the Categorical Exploratory Data Analysis (CEDA) computing paradigm to extract all potential pieces of associative information, and then employs Kolmogorov's randomness-proper concept to confirm and select those reliable ones. All confirmed pieces of associative information across the spectrum of Re-Co dynamics collectively constitute the major part of Class-informatics under the Taxonomic Hierarchy.

Specifically, CEDA exclusively operates on the contingency table platform within a fully categorical or categorized Re-Co dynamics to extract all potential pieces of associative information in a form of readable, visible and explainable 1D histogram. When categories of the response variable are arranged along the column-axis of the contingency table, for instance, then such a 1D histogram is a row-vector of contingency table. Its meaning is read, visualized and explained as being a finite approximation of conditional distribution of response variable given a category of a covariate feature or feature-set. Therefore, the randomness pertaining to any contingency table is explicitly identified as a collection of column-wise Multinomial randomness conditioning on the vector of column-sums, which stands for the sample marginal distribution of the response variable.

With known and explicit randomness of a contingency table, each computed piece of associative information is coupled with its idiosyncratic finite sample precision based on alternative and null ensembles of simulated contingency tables that are simulated according to Kolmogorov's randomness-proper concept. So that, each potential piece of associative information has its own individual finite sample precision, with which its reliability check is performed. Each confirmed piece of associative information is called a major feature-category of $k$-order with $k$ being the dimensionality of covariate feature-set.

\paragraph{Regarding heatmaps:}
It is reiterated that all confirmed and selected major feature-categories of various orders are collectively displayed in a heatmap, which is a format of bipartite network that reveals the pattern-formations between involved study-subjects and selected major feature-categories. With all study-subjects and major feature-categories are arranged along the column- and row-axes, respectively, a heatmap as a binary bipartite network of subject-memberships of all selected major feature-categories of various orders.  When permuting along the row- and column-axes via two HC-trees, its block-structures are framed to consequently reveal vertical block-chains that map out recognizable associative relational linkages from topological neighborhoods of annotated subject-space to clusters of major feature-categories, while horizontal block-chains map out mechanistic dependence in reverse neighborhood correspondences.

That is, topological similarity among involved study subjects becomes visible with multiscale clustering structures, while topological similarity among selected major feature-categories brings out all kinds mechanistic dependence, which are waiting to be explained and explored. As such a heatmap is adopted as the primary platform for exhibiting Re-Co dynamics.  As such the most striking character defining Class-informatics is that each response-category annotated neighborhood of study-subjects' individual character-landscape is embedded with a serial mechanistic dependence revealing topological neighborhoods of major feature-categories.

When a Re-Co dynamics defined by a response-variable with categories being just a subset of the collection of all classes,
An extra merit of such a heatmap is that it also provides another source of information regarding each non-participating study-subject's individual geometry-outlier and information-outlier statues. Both outlier status and associative information pieces indeed can be pooled together, organized and displayed through one single heatmap, as would demonstrated in later section. This heatmap in fact provide a rigorous resolution for outlier detection. That is, this heatmap resolves the critical outlier detection question mentioned in Tukey's 1962 paper.

\paragraph{Regarding taxonomy sustained by Class-informatics:} In view of contexture details, behind taxonomic and classification tasks, information content of Class-information contained in data has two fronts. The first front is on each study subject's many versions of individual character-landscapes. Each version is correspondingly embedded within all extracted pieces of associative information with respect to a specific categorical response variable. By properly organizing such intertwined information pieces, all classes are collectively and fully described with informative details. This spectrum of information content is the basis for motivating class-specific knowledge. This front of Class-informatics should have been advocated in all scientific literatures.

The second front is regarding heterogeneous mechanistic dependence collectively reveal through computed and clustered associative information pieces. This front of mechanistic dependence provides the vital ingredients for revealing underlying dynamics of complex system of interest. These two fronts are to be explicitly demonstrated through heatmap's block-structures to jointly facilitate better understanding among all involved classes. As such the foundation of good explanations is built by a computational taxonomist. Therefore, collectively Class-informatics provides ``good explanations'' regarding the underlying dynamics of complex system. All inferential decision-makings are necessarily made in accord with such Class-informatics and explicitly explained.

\section{Technical setups for SDA in a complex system.}
When facing a complex system giving rise to phenomena of scientific interest and importance, scientists instinctively want to gain some understanding ``what is out there and how it works''. Consider a very simplistic complex system with a categorical response variable ${\cal Y}$ taking ``values'' from a finite set of class-IDs, say $\{ID[1], ID[2], ID[3],..., ID[l],.., ID[L]\}$. To be able to see this system better as generic system scientists, taxonomists will build an intrinsic hierarchy, such as tree-geometry, upon this collection of class-IDs. This multiscale structure is not unusual in real-world because domain knowledge typically clusters among classes regarding: ``which is closer to which, but far away from which''. At the beginning of data analysis, taxonomists as data analysts do not need to explicitly know priori about this hierarchical structure. The key point here is that such structure would be naturally emerged out of the data-driven Class-informatics. For the meantime, we denote this hierarchical geometry as ${\cal T}[{\cal Y}]$.

For studying such a complex system represented by ${\cal Y}$, the data curators or subject matter scientists must also select a collection of covariate features or variables denoted as ${\cal X}=\{X_1, X_2,..., X_K\}$. For simplicity, but without loss of generality here, each $X_k,\; k=1,.., K$ is taken to be 1D feature of measurement or observation of quantitative data type. It is understood that ${\cal X}$ has been selected by the scientist or data curator according to his/her domain knowledge and experiences. Such a selection act is equivalent to an encoding act of domain knowledge and experience. This behind-the-scenes encoding indeed is one of the most essential recognition when analyzing a data set pertaining to a targeted complex system of scientific interest. Since this covariate feature-set ${\cal X}$ indeed defines the scope of explorations into the targeted complex system. Via such explorations, ideally, the scientists intend to reveal and explain the original system's underlying dynamics through the directional associative relationships from ${\cal X}$ to many versions of subsets of ${\cal Y}$, says ${\cal Y}_{sub}$, in a form of branch-vs-branch comparisons defined through the so-called Taxonomic Hierarchy ${\cal T}[{\cal Y}]$. In this sense, ${\cal T}[{\cal Y}]$ must somehow captures the original dynamics underlying the complex system under study.

It is surprising that such exploratory efforts of Taxonomy might results pattern information far beyond the domain scientist or data curator's understanding about the system. In fact, what is ``beyond'' is known to be high order interacting effects. They are the exact key and critical information that has gone missing in almost all literatures involving data analysis. Since under the modeling-prediction doctrine, all adopted models just can't accommodate interacting effects, even for order-2, effectively and precisely. In contrast, SDA is capable of handling, exploring and carrying out all orders of interacting effects within ${\cal X}$.

As such a scientific inquiry of taxonomy nature is translated and defined by ``better and full understanding the associative relationships from ${\cal X}$ to a spectrum of ${\cal Y}_{sub}$ specified by ${\cal T}[{\cal Y}]$ '', where ${\cal Y}_{sub}=\{ID[l_1],...., ID[l_m]\}\subseteq \{ID[1], ID[2], ID[3],..., ID[l],.., ID[L]\}$ with $L \geq m \geq 1$. It is evident that ${\cal Y}_{sub}$ only involves with data-points annotated by one of IDs in $\{ID[l_1],...., ID[l_m]\}$. We can see the differences between  ${\cal Y}$  and  ${\cal Y}_{sub}$ through their scatter plots with respect to any generic covariate-feature subset: ${\cal X}_{sub}=\{X_{k_1}, X_{k_2},..., X_{k_h}\}\subseteq {\cal X}$ with $K \geq h \geq 1$ pertaining to ${\cal Y}$ and ${\cal Y}_{sub}$. The geometry of ${\cal Y}_{sub}$ would have many ``holes'' comparing with the one of ${\cal Y}$ because of missing all data points belonging to the complement subset ${\cal Y}^{(C)}_{sub}={\cal Y}-{\cal Y}_{sub}$. See multiple Iris example illustrations in Fig~\ref{pairplot}. More details are given below. This is true for all possible ${\cal X}_{sub}$. Hence, it is obvious that the classification Signal-to-Noise (S-N) ratios across all possible ${\cal Y}_{sub}$ vary significantly. This is the first clue of hierarchy-based information complexity embedded within Class-informatics.

The two definitions of ${\cal X}_{sub}$ and ${\cal Y}_{sub}$ are unrelated. In this paper, we emphasize and advocate the fact that, contrasting information contents of any specific Re-Co dynamics: ${\cal X}$ to ${\cal Y}$ or ${\cal X}$ to ${\cal Y}_{sub}$, respectively, would involve with all possible choices ${\cal X}_{sub}$. Within the specific Re-Co dynamics: ${\cal X}$ to ${\cal Y}$, a choice of ${\cal X}_{sub}$ will give rise to a $h$-dim point-cloud geometry involving all data-points annotated by members of ${\cal Y}$, while, within the Re-Co dynamics: ${\cal X}$ to ${\cal Y}_{sub}$, the same ${\cal X}_{sub}$ would give rise to another $h$-dim point-cloud geometry involving only a subset of data-points annotated by members of ${\cal Y}_{sub}$. These two $h$-dim point-cloud geometries can be highly distinct at least from two aspects. The first aspect is due to missing data points annotated by members of complement subset ${\cal Y}^{(C)}_{sub}={\cal Y}-{\cal Y}_{sub}$. The 2nd aspect is due to the potential outlier statuses of data-points of complement subset ${\cal Y}^{(C)}_{sub}$. This 2nd aspect has hardly even been studied rigorously and reported explicitly in literature. Nevertheless, Re-Co dynamics: ${\cal X}$ to ${\cal Y}$ and ${\cal X}$ to ${\cal Y}_{sub}$ supposedly are closely related. A glimpse of such relatedness would be seen through a contingency table platform as follows.

For expositional simplicity, consider the case of all 1D features in ${\cal X}_{sub}$ being already categorized. The associative relations from the $h$-dim point-cloud geometry of ${\cal X}_{sub}$ to ${\cal Y}$ is fully contained in a hyper-contingency table of $h$-dimension, denoted as $HCT[{\cal X}_{sub}; {\cal Y}]$, where all categories partitioning the $h$-dim point-cloud geometry of ${\cal X}_{sub}$ are arranged along the row-axis, while categories of $\{ID[1], ID[2], ID[3],..., ID[l],.., ID[L]\}$ arranged along the column-axis. Likewise, a contingency table for associative relations from ${\cal X}_{sub}$ to ${\cal Y}_{sub}$ is contained in the contingency table $HCT[{\cal X}_{sub}; {\cal Y}_{sub}]$. Specifically speaking, pieces of associative relations contained in $HCT[{\cal X}_{sub}; {\cal Y}]$ and $HCT[{\cal X}_{sub}; {\cal Y}_{sub}]$ are both manifested through their row-vectors, individually. They are empirical conditional distributions of ${\cal Y}$ and ${\cal Y}_{sub}$ given a category of ${\cal X}_{sub}$, respectively. The later might have surprisingly high S-N ratio, while the former one might not. Further, the two versions of point-cloud-geometries of ${\cal X}_{sub}$ under ${\cal Y}_{sub}$ and ${\cal Y}^{(C)}_{sub}$, respectively and collectively reveal geometric outlier information. As such the concept of ``Class-informatics'' are further explained as follows.

Our explanation rests on the key point that the above two kinds of associative relations in fact are linked by the information of ``outlier'' in the following fashion. Let $ID[l^*] \in {\cal Y}^{(C)}_{sub}$. Consider the scenario that the covariate part of a data-point of $\{{\cal X}_{sub}; {\cal Y}^{(C)}_{sub}=ID[l^*]\}$, say ${\cal x}^*_{sub}$, is confirmed as an ``outlier'' with respect to not only the physical $h$-dim point-cloud geometry (PCG) of ${\cal X}_{sub}$, denoted as $PCG[{\cal X}_{sub}|{\cal Y}_{sub}]$, but also from the to-be-defined information perspective of an ``outlier'', see section below. Then, this confirmed piece of information of ``outlier'' status indeed immediately points to the fact that the to-be-annotated response categories for ${\cal x}^*_{sub}$ must be one member of  ${\cal Y}^{(C)}_{sub}$. As such, this confirmed piece of outlier information is very valuable.

Given that there are so many ${\cal X}_{sub}$ out of ${\cal X}$ and ${\cal Y}_{sub}$ out of ${\cal Y}$, the collective of all such pieces of information should be very critically shedding lights on hierarchical-geometry ${\cal T}[{\cal Y}]$ for the task of taxonomy and consequently for performing the task of Classification, as would be demonstrated vividly in this paper through the Iris example. This collective of all such kinds of information is the foundation of ``Class-informatics''. Nonetheless, such information is supposed to be provided by any knowledgeable taxonomist of such a complex system.

In summary, from this SDA perspective, the nature of taxonomy and classification assisted by the outlier detection is indeed systemic. Further, a brand-new way of perceiving the issue of outlier detection, as being systemic by involving not only point-cloud geometric, but also information perspectives.

  \begin{figure}[h!]
 \centering
\includegraphics[width=1.0\textwidth]{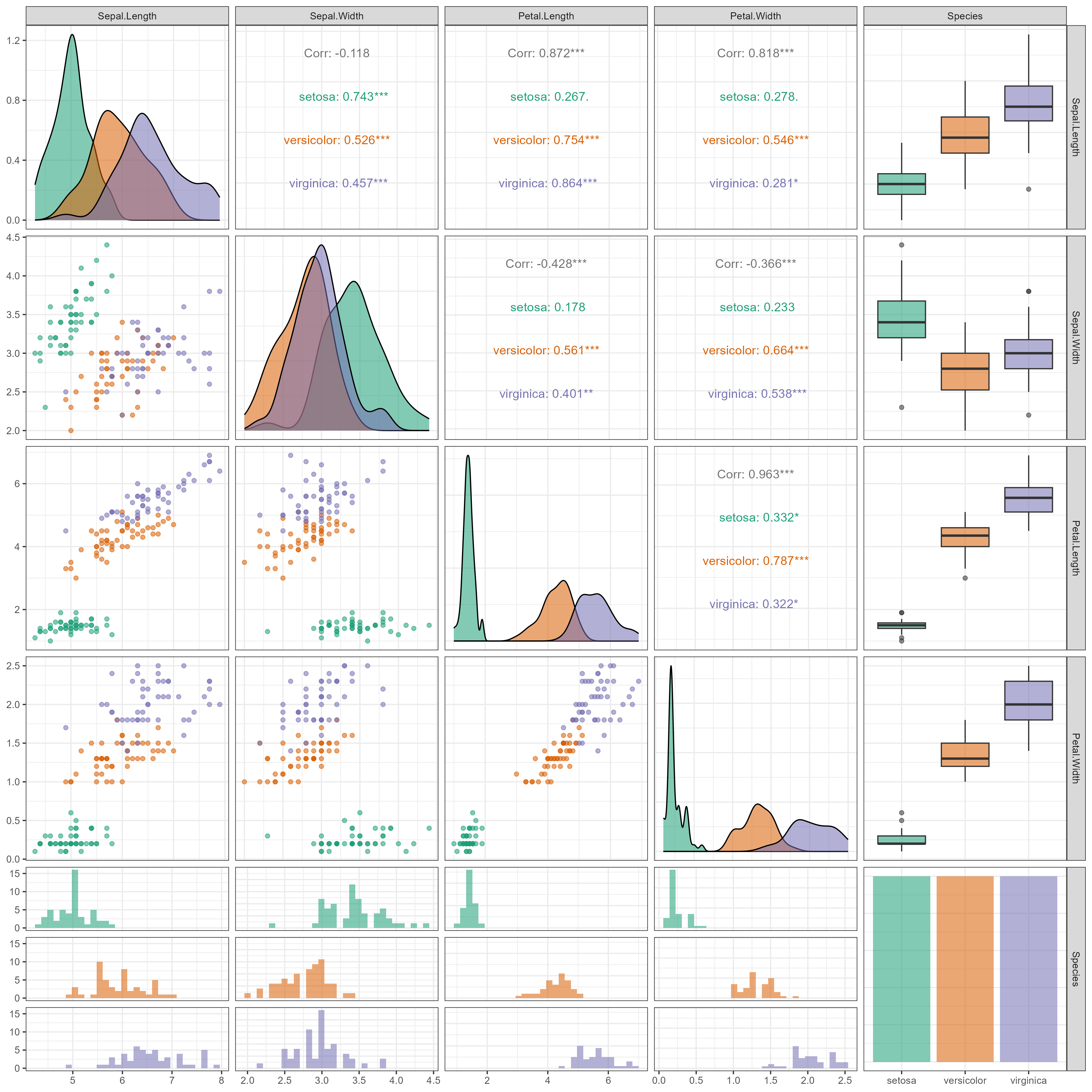}
 \caption{Pairplot of Iris data. }
\label{pairplot}
 \end{figure}

\section{Iris example}
Here, we realistically illustrate Class-informatics through a simple, but real example: Iris data. Through multiple glimpses of this data set, some clues of its Class-informatics can be seen. All $4\times 4$ array-panels in Fig~\ref{pairplot} respectively reveal six distinct point-cloud geometries pertaining to six covariate feature-pairs of Iris-IDs: ${setosa, versicolor, virginica}$, encoded by a color-coding scheme. When dropping all data-points belonging to anyone of three Iris-IDs, we would create point-cloud geometries that visually highlight all individual roles of missing data-points in the point-cloud geometry of full data set. Geometrically speaking, some individual data-points evidently signal their outlier status, while some are difficult to notice. Any outlier status is a piece of strong signal of its Iris-ID. This pure geometric signal is one-data-point's Iris-ID-revealing idiosyncratic characteristic. It is fundamentally distinct to any associative patterns of collective nature. That is, this signal might not be captured by Re-Co dynamics of ${\cal X}$-vs-${\cal Y}$. This is one locality-specific aspect of Class-informatics of fine scale.

On the other hand, Class-informatics of large scale can be seen through the dependence structures of any feature-pairs. Surely they vary significantly from subsystem-to-subsystem. For the subsystem: ${\cal Y}_{sub}=\{versicolor, virginica\}$, many feature-pairs provide evidently heterogeneous dependence patterns, while in contrast with the subsystem: ${\cal Y}_{sub}=\{setosa, virginica\}$, all feature-categories order-2 can capture patterns with exclusive differentiating capability. Thus, this is another aspect of Class-informatics of another scale. This aspect also points to the varying S-N ratios across all three subsystems.

In particular, it is evident that point-cloud geometries of all feature-pairs under the subsystem: ${\cal Y}_{sub}=\{Setosa, Virginica\}$, are equipped with strong outlier-detecting capability due to the apparent separation between Setosa's point-cloud and Viginica's. Such a capability should be helpful for classifying data-points belonging to $versicolor$. In other words, the ${\cal Y}_{sub}=\{Setosa, Virginica\}$ subsystem's dependence patterns become clearer and specific, as such the S-N ratio likely become larger. That is, signal strengths of majority of feature-categories become stronger. In comparison, the S-N ratios under subsystem ${\cal Y}_{sub}=\{versicolor, virginica\}$ in general are not as strong. This fact also implies the pair $\{versicolor, virginica\}$ is closer than the pair $\{Setosa, Virginica\}$.

Next, we see how the fine-scale and large-scale of Class-informatics interacts. Each of such large S-N-ratio feature-pairs would turn its point-cloud geometry into a large-scale basis for detecting outsiders of this subsystem. This outlier detection is necessary when carrying out during the process of categorization on the quantitative feature-pair. And this declaration of outlier is “once for all”. It is noted that such fine-and-large scales interacting geometric information pieces are ignored by all approaches that only focus on associative relationships. For instance, modeling-based methodologies have limited capability of capturing such kind of information. On the other hand, in machine learning, even the Random Forest and variants of boosting algorithms will not pick up such a kind of information due to the limits of decision-tree. Such limits can be traced to decision-tree’s incapability of dealing with interacting effects via irregular geometric shapes of manifold of any orders. Thus, such limits are attributed to the known fact that Random Forest and boosting algorithms suffer large error rates when facing large number of response-IDs. This known phenomenon is in particular severe under so-called imbalanced settings when some classes have much larger sample sizes than other classes.

As would be demonstrated in sections in the rest of this paper, Class-informatics of Iris data will include all possible feature-sets of all orders across all possible subsystems. Its full information content in data would include the collection of pieces of associative information and outlier statuses of study-subjects. With Class-informatics in hand, the data analysts can answer questions regarding the original complex system and give good explanations like a taxonomist does. This is why we can advocate the concept of ``creating a virtual knowledgeable Taxonomist'' in this paper.

\section{Scientific Data Analysis (SDA): Full system illustrated.}
Via the generic data representation discussed previously, Iris data set naturally has such a categorical response (Re) variable:
 \[
 {\cal Y}\in \{ID[1], ID[2], ID[3]\}=\{setosa(Se), versicolor(Ve), virginica(Vi)\}
 \]
 and four continuous covariate (Co) feature variables:
 \[
 {\cal X}=\{X_1, X_2, X_3, X_4\}=\{Sepal length(SL), Sepal width(SW), Petal length(PL), Peta width(PW)\}.
 \].
The sample size of 4-dim quantitative data points annotated by each one of the three response-IDs of ${\cal Y}$ is 50. So, the total sample size $n$ is 150. In this section, we illustrate all computational protocols in SDA and all kinds of fabrics of Iris' Class-informatics.

Such readable, visible and explainable information content in Iris' Class-informatics is supposed to include full spectrum of associative patterns from ${\cal X}$ to ${\cal Y}$ and to all versions of ${\cal Y}_{sub}$. As such Iris' Class-informatics also includes outlier statuses of all data points of ${\cal Y}^{(C)}_{sub}$ when any Re-Co dynamics from ${\cal X}$ to ${\cal Y}_{sub}$ is analyzed. It is worthwhile to reiterate that the outlier detection task and the categorizing task are two sides of one coin because all covariate features here are quantitative. Specifically, a data point's ${\cal X}_{sub}$ component-vector can only be properly categorized if it must not be a ``geometric outlier'' with respect to ${\cal X}_{sub}$ manifold constituted by data points annotated by response-IDs in ${\cal Y}_{sub}$. Further, after all outlier detection (OD) tasks with respect to all selected ${\cal X}_{sub}$, its collection of categories will then afford checking whether it is an ``information outlier'' or it indeed can be suitably annotated in a predictive fashion.

\begin{figure}[h!]
 \centering
\includegraphics[width=1.0\textwidth]{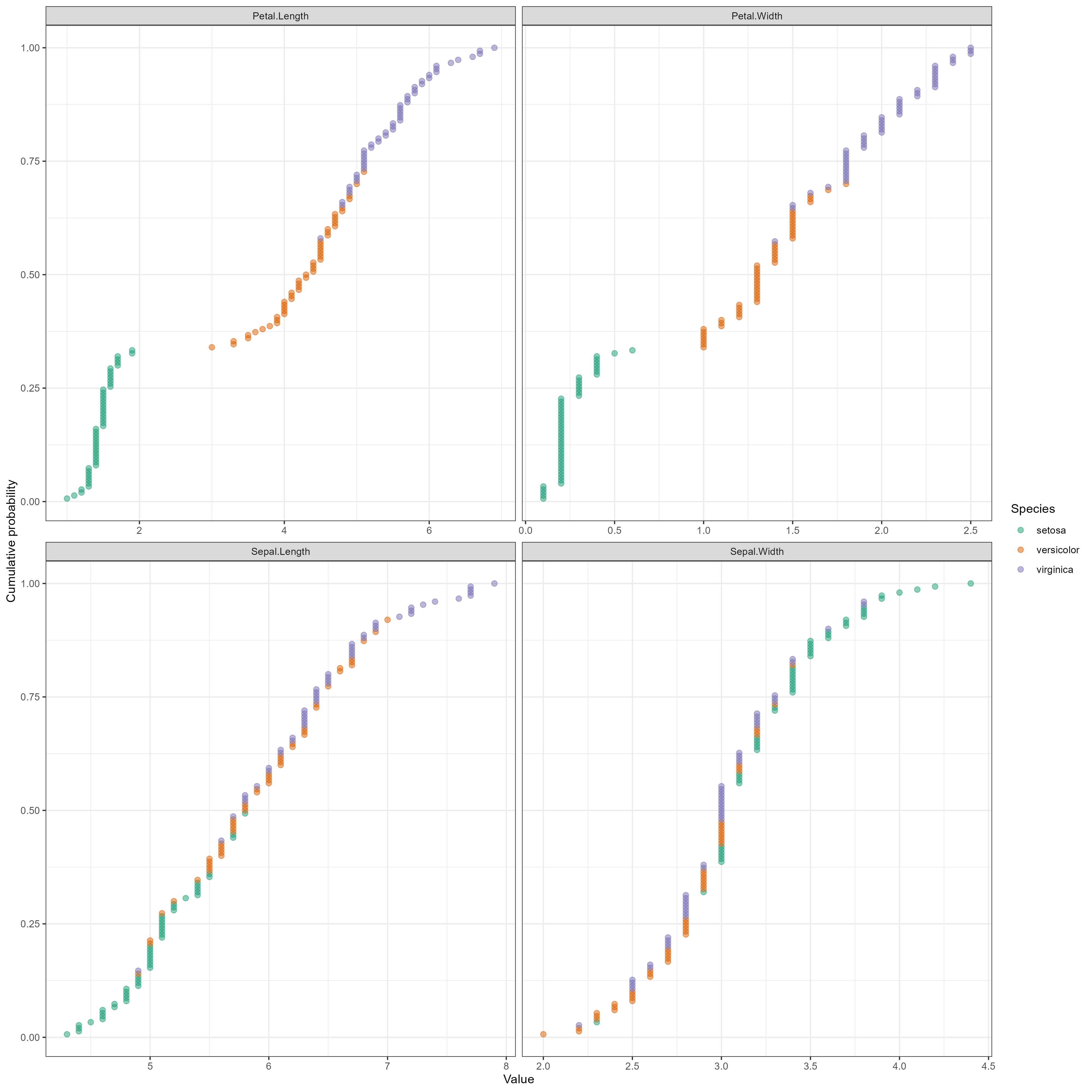}
 \caption{Empirical cumulative distribution function (ECDF) annotated with three species-IDs of four covarite features in Iris data.}
\label{cdfplot}
 \end{figure}

SDA begins with looking at the randomness pertaining to the 1D ensemble of 150 data points of each feature variable of ${\cal X}$. The targeted characters are two kinds of complexity: one is embedded within this randomness and another one is imbedded within interacting relations between this randomness and annotated response-IDs of ${\cal Y}$. Though these two kinds of randomness related complexity are not well-defined here, but they are indeed visible. The first kind of complexity is displayed through the empirical cumulative distribution function (ECDF) of $\{X_1[i]|i=1,.., 150\}$, while the second kind can be seen through the species-colored-encoded ECDF of $(X_k[i], Y[i])$ with $k=1, 2, 3,4$, as respectively shown in the panels (A) to (D) in Fig~\ref{cdfplot}.

Upon ECDF in all four panels, we see the randomness pertaining to the 150 measurements of $X_k$ in a highly spreading fashion. Such visible randomness is hard-to-describe because of lacking any aggregating traces. That is, the complexity of ECDF is in our way of achieving any concise description of such randomness. Each species-colored-encoded ECDF in the four panels is a fully sufficient statistics because it can regenerate the observed ensemble $\{(X_k[i], Y[i])|i=1,.., 150; k=1, 2, 3,4\}$. So, it is supposed to carry all possible associative patterns. However, again it is the complexity of ECDF is in our way of describing any of associative patterns. If we further smooth the ECDF to derive a smooth density estimation, then the two kind of complexity are increased, not reduced.

\begin{figure}[h!]
 \centering
\includegraphics[width=1.0\textwidth]{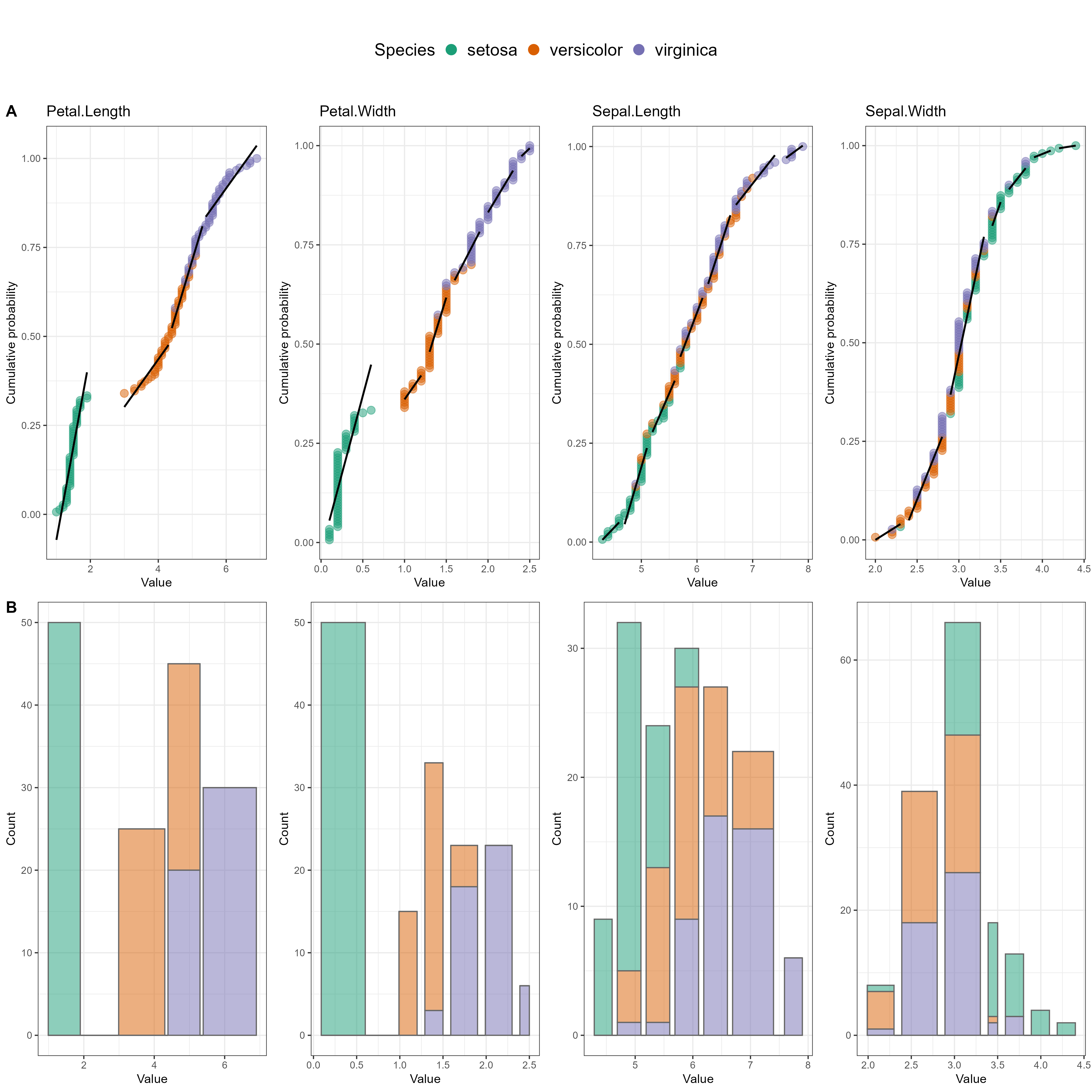}
 \caption{(A) ECDF and (B) histogram of four features of Iris data. }
\label{Irishistogram}
 \end{figure}

Therefore, we need to reduce the complexity of ECDF in order to see $X_k$'s randomness and associative patterns from $X_k$ to ${\cal Y}$. This task is done and presented in Fig~\ref{Irishistogram}. Here the complexity reduction on each of the four ECDFs of $X_k$ with $k=1,.., 4$ is done by a piecewise linear approximation to it \cite{FR2018}. Such a piecewise linear approximation in Fig~\ref{Irishistogram} is a Hierarchical Clustering (HC) based solution to an optimization problem. Each piecewise linear approximation to an ECDF corresponds to a histogram. Each of its bar is color-encoded with numbers of members from the three Iris species to reveal bar-specific associative pattern $X_k$ to ${\cal Y}$. So pattern of randomness of $X_k$ and its associative relations with ${\cal Y}$ become visible. Thus, {\bf any formation of readable, visible and explainable information from data-points need some kinds of aggregating mechanisms to reduce the seemingly abstract, but obviously inherent complexity in all data-representations.} It is critical to note that histogram allows us to make concise {\bf ``shape'' comparisons} among species-specific distributions.

By constructing histogram to aggregating similar data-points into bins, the price for such visible randomness is indeed rather low. Since each histogram is a basis of mimicking the observed data of $X_k$ according to the underlying reasoning given as follows. As a histogram's distribution is a piecewise linear approximation to an empirical distribution function that is supposedly embraces the whole spectrum of randomness pertaining to an 1D covariate feature. That is, a histogram preserves nearly the entire regime of randomness contained in the whole set of 1D measurements. In particular, each bar or bin of a histogram is equipped with uniform randomness with two locality characteristics: Width and count. The width stands for variation and counts for intensity. The locality's variations are further conditioning out because all three species' histograms share the same set of bins. In other words, these three histograms have been standardized for ``shape'' comparisons. Consequently, locality's intensity would be equal to the sum of 3-dim vector of species-specific counts.

Each bin-locality's 3-dim vector of counts becomes one (column) vector of this contingency table. That is, we can transform a species-color-encoded histogram into an observed contingency table: all bins are arranged along the column-axis and each bin's counts are respectively separated as counts belonging to the three Iris species arranged along the row-axis. A version of species-color-encoded histograms of $X_k$ is turned into a contingency table as shown in Table~\ref{tab:Iristable0}. It is essential to note that each contingency table is a basis for comparing $X_k$-category specific conditional distributions of ${\cal Y}$. That is, such a contingency table is how nearly all randomness patterns and associative relations are preserved after conditioning on bin-boundary information. This contingency table platform lays the foundation for carrying out Categorical Explanatory Data Analysis (CEDA) of SDA computing.

\begin{table}[t]
\centering
\caption{Contingency table of Iris species across bin IDs for the encoded features.}
\label{tab:Iristable0}
\fontsize{8pt}{9pt}\selectfont
\begin{adjustbox}{width=\textwidth}
\begin{tabular}{lrrrrrrrrrrrrrrrrrrrrrrrr}
\toprule
 & \multicolumn{4}{c}{Petal.Length}
 & \multicolumn{6}{c}{Petal.Width}
 & \multicolumn{7}{c}{Sepal.Length}
 & \multicolumn{7}{c}{Sepal.Width} \\

\cmidrule(lr){2-5}
\cmidrule(lr){6-11}
\cmidrule(lr){12-18}
\cmidrule(lr){19-25}

Species
& Bin \#1 & Bin \#2 & Bin \#3 & Bin \#4
& Bin \#1 & Bin \#2 & Bin \#3 & Bin \#4 & Bin \#5 & Bin \#6
& Bin \#1 & Bin \#2 & Bin \#3 & Bin \#4 & Bin \#5 & Bin \#6 & Bin \#7
& Bin \#1 & Bin \#2 & Bin \#3 & Bin \#4 & Bin \#5 & Bin \#6 & Bin \#7 \\

\midrule
setosa
& 50 & 0 & 0 & 0
& 50 & 0 & 0 & 0 & 0 & 0
& 9 & 27 & 11 & 3 & 0 & 0 & 0
& 1 & 0 & 18 & 15 & 10 & 4 & 2 \\

versicolor
& 0 & 25 & 25 & 0
& 0 & 15 & 30 & 5 & 0 & 0
& 0 & 4 & 12 & 18 & 10 & 6 & 0
& 6 & 21 & 22 & 1 & 0 & 0 & 0 \\

virginica
& 0 & 0 & 20 & 30
& 0 & 0 & 3 & 18 & 23 & 6
& 0 & 1 & 1 & 9 & 17 & 16 & 6
& 1 & 18 & 26 & 2 & 3 & 0 & 0 \\

\bottomrule
\end{tabular}
\end{adjustbox}
\end{table}

\subsection{Finite sample precision through Kolmogorov's randomness-proper.}
The next foundational step of SDA computing is to recognize the randomness embraced by each of the four contingency tables in Table~\ref{tab:Iristable0}. This recognition of randomness in contingency table is prepared for evaluation of finite sample precision for each associative relational information from each category of $X_k$ to ${\cal Y}$. This evaluation is advocated by following Kolmogorov's randomness-proper concept\cite{kolmogorov}. To implement this concept in data analysis, we propose to manifest the fully recognized randomness embraced within a contingency table by constructing two kinds of ensembles of simulated contingency tables: Observed (or Alternative) and null.

For the alternative ensemble, the observed randomness recognition is described as follows. The vector of column-sums stands for the vector of locality intensity approximating its ECDF of the marginal distributional randomness of $X_k$. In contrast each row-vector stands for an approximation of the conditional distribution of $X_k$ given a category of ${\cal Y}$, which is a class-ID, with a fixed row-sum 50. This is observed intrinsic randomness contained within a contingency table of $X_k$-vs-${\cal Y}$. Therefore, it is natural to make use of Multinomial randomness to capture the randomness pertaining to row-vector of counts. As such, by constructing a contingency table in row-by-row fashion, a mimicry of the observed contingency table is resulted. Such concepts and operations become the foundation for mimicking observed contingency tables for the alternative ensemble.

As for the null ensemble of contingency table, we generate a pseudo-$X_k$ (denoted as $X^*_k$) with marginal distributional randomness specified by the vector of column-sum proportions. $X^*$ has the same marginal distributional randomness as $X_k$, but is independent of ${\cal Y}$. Given each row-sum being fixed at 50, we simulate each species-specific row-vector by making use of Multinomial randomness with the randomness pertaining to the marginal distributional randomness of $X_k$. In this fashion, we construct a contingency table of a $X^*_k$-vs-${\cal Y}$, which is called a null contingency table. As such this $X^*_k$ has absolutely zero association with ${\cal Y}$. This is how the null ensemble of null contingency tables is built.

\begin{figure}[h!]
 \centering
\includegraphics[width=1.0\textwidth]{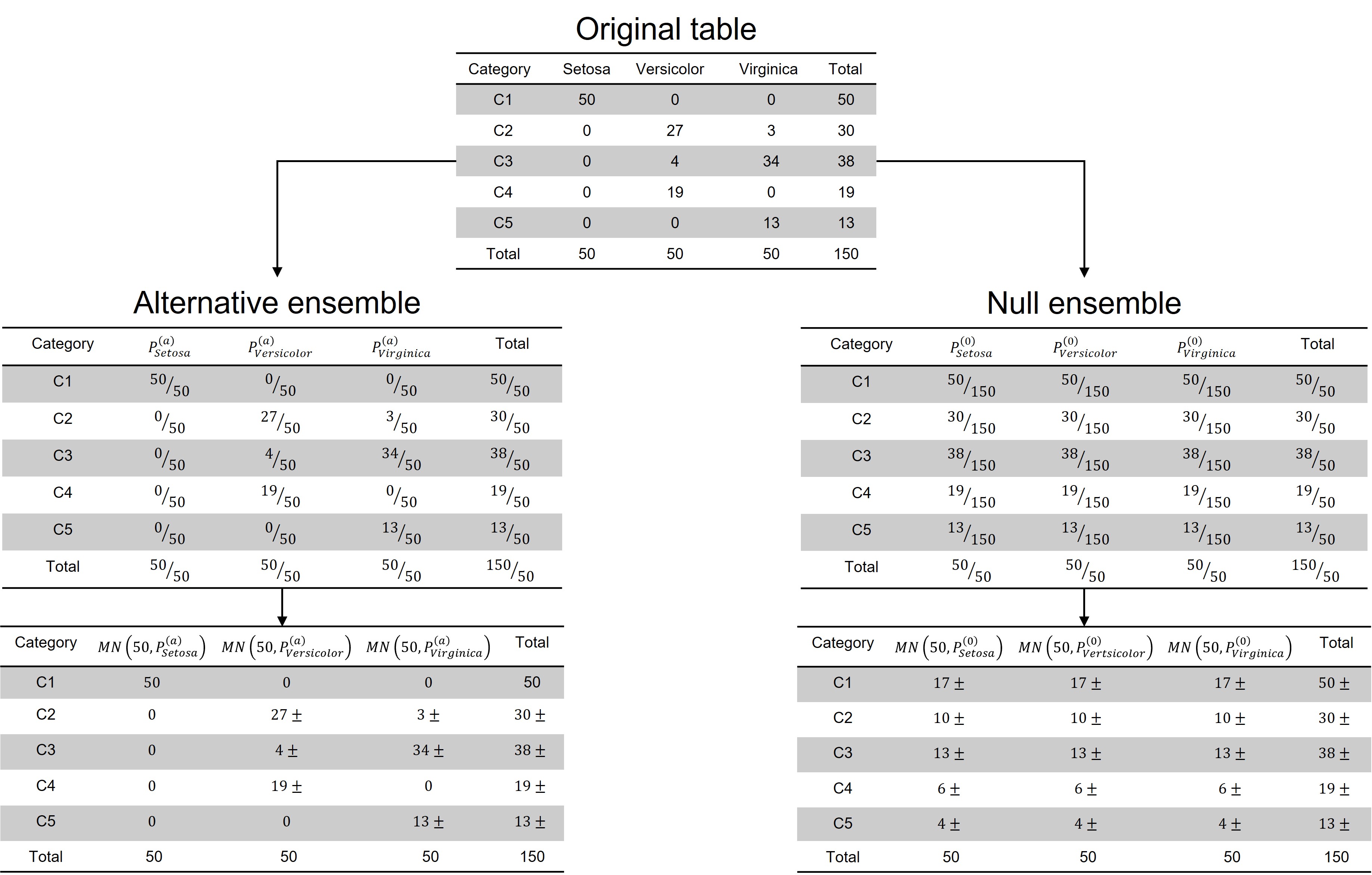}
  \caption{Alternative and null ensembles for randomness of the observed contingency-table of Petal length in Iris data. }
\label{summary}
 \end{figure}

As alternative ensemble of mimicries of observed contingency tables retain the associative pattern information almost completely, while a null ensemble of null contingency tables supposedly destroy all observable associative patterns, the following five key points of constructing this pair of ensembles are critically important in SDA computing. These five key points are narrated according to a $5\times 3$ contingency table of $X_3-vs-{\cal Y}$ as displayed in the top panel of Fig~\ref{summary}, which 5 categories: $(C1, C2,.., C5)$, of Petal length are arranged along the row-axis and the three species as three response-categories on column-axis.
\begin{description}
\item[A.] Equally distributed finite sample sizes for all corresponding vectors of row-sums of both alternative and null contingency tables.
\item[B.] Each column-vector of any null contingency table is distributed centered around the observed vector of row-sum-proportions, which is $(\frac{50}{150}, \frac{30}{150}, \frac{38}{150}, \frac{19}{150}, \frac{13}{150})$. See the right table of the 2nd panel-layer of Fig~\ref{summary}.
\item[C.] In equal finite sample sense, any row vector of alternative contingency table standing for an 1D histogram approximation of conditional distribution of ${\cal Y}$ given a category of $X_k$ is comparable with the corresponding row vector of null contingency table standing for an 1D histogram approximation of conditional distribution of ${\cal Y}$ given the same category of $X^*_k$. This comparison is seen via the two tables: Left one for alternative ensemble and the right one for null ensemble on the 2nd panel-layer of Fig~\ref{summary}.
\item[D.] This comparison with equal sample size is the foundation for the finite sample precision of any piece of associative relational information from a category of $X_k$ to ${\cal Y}$. See an illustrative comparison on the 3rd panel-layer of Fig~\ref{summary}.
\item[E.] This basis of finite sample precision will facilitate a reliability check for each candidate piece of associative relation information.
\end{description}

Nowadays, carrying out HC algorithm to build a HC-tree has been a simple task. Hence, for each of $X_k\; k=1,.., 4$, it is potential and feasible to build data-driven HC-tree based contingency table to offer candidates of informative associative patterns. Such operations will also be carried out for pairwise, triplet-wise, quartet-wise,..etc. quantitative 1D features for exploring candidate pieces of associative relational information via interacting effects of order-2, order-3, order-4,..etc. As such {\bf SDA can explore interacting effects of any orders via HC-tree based contingency table platform.} This is a capability not seen in Statistics or Machine Leaning and other literature.

According the above listed five key points, any contingency table will be coupled with a pair of its idiosyncratic ensembles of simulated contingency tables to facilitate finite sample precisions for all computed candidate pieces of associative relational information. Hence, each candidate piece of associative relational information will be checked regarding its idiosyncratic reliability with respect to a threshold of scientist's choice, as would be discussed below.

Here we narrate the explicit meaning of associative pattern information revealed through contingency table platform. Theoretically speaking, using the format of contingency tables as illustrated in Fig~\ref{summary}, the content of a piece of associative relational information is exhibited through the comparison of a row-specific vector of proportions with the vector of column-sum-proportions. The larger degree of distinction between these two vectors of proportions is, the stronger signal of associative pattern is exerted by this row-vector, or by the locality of a category of $X_k$. Such a comparison can be carried out through the difference of their Shannon entropies. In this Iris example, the vector of column-sum-proportions of all contingency tables, like that in Fig~\ref{summary}, is $(1/3, 1/3, 1/3)$. This column-sum-proportion vector indeed achieves the largest Shannon entropy value $\log 3$. Further, since Shannon measures the degree of uncertainty, this property makes the task of explaining computational results straight forward. In this Iris example, the strength of a piece of associative relational information pertaining to a category of $X_k$ is evaluated via how much smaller is its row-specific entropy than $\log 3$. This is a Shannon entropy based natural evaluation of signal strength with clear meaning. As clearly stated in the above listed statement-[A], this signal strength must reflects the sample size pertaining the involving category of $X_k$.

\begin{figure}[h!]
 \centering
\includegraphics[width=1.0\textwidth]{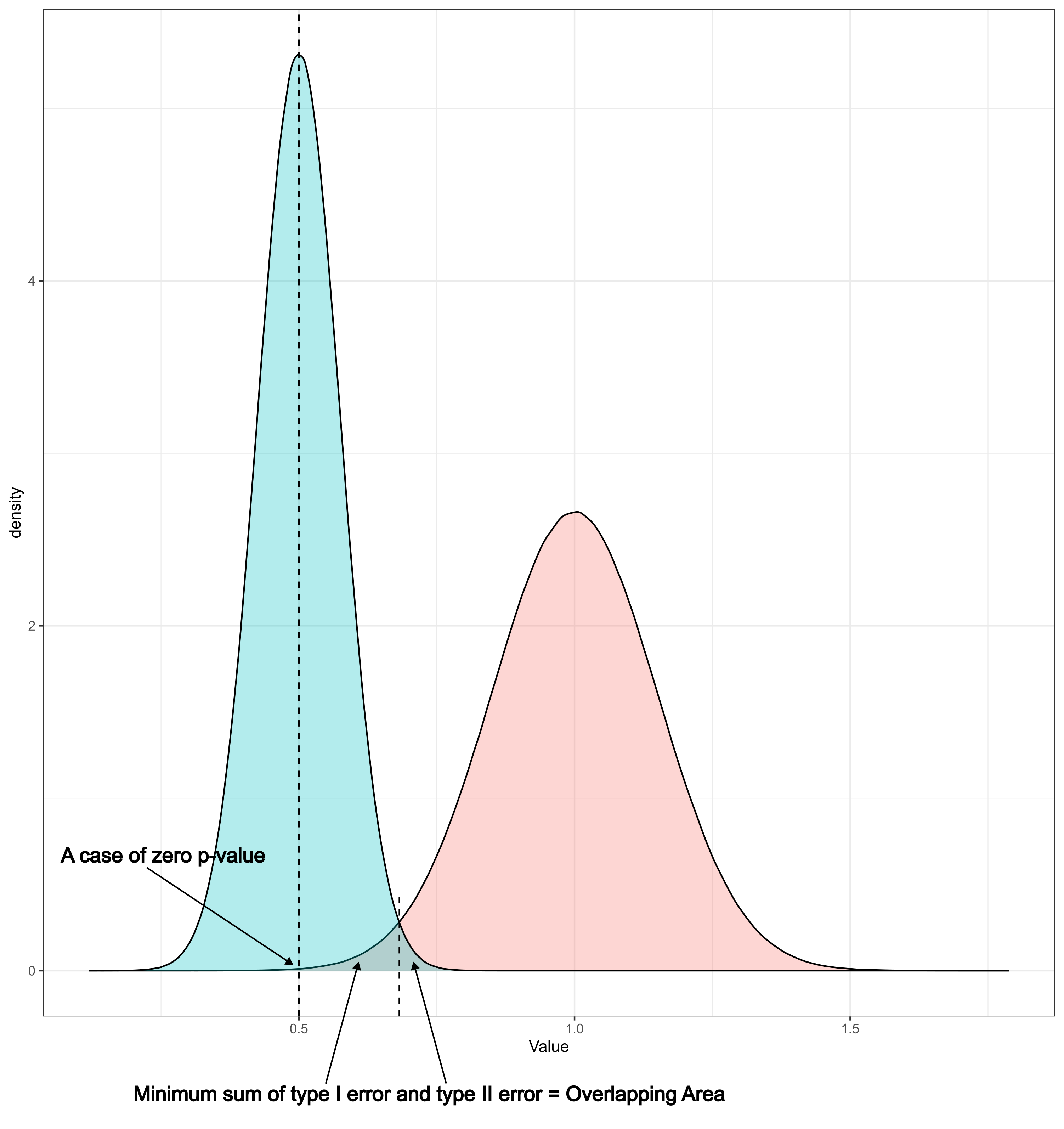}
\caption{Illustrative figure of alternative-vs-null entropy distributions with marked Type-I and Type-II errors in comparison with p-value. }
\label{alternullok}
 \end{figure}

As aforementioned that a reliability check is needed for any associative pattern information. By having explicit meaning for a signal strength of associative pattern information pertaining to a category $X_k$ via Shannon entropy based comparison, its idiosyncratic reliability check must depend on its idiosyncratic pair of alternative and null entropy distributions as explicitly laid out through statements-[C] to [E] as illustrated in Fig~\ref{alternullok}. In this figure, we clearly see the following essential messages emerging out as follows.
\begin{description}
\item[a.] The alternative and null entropy distributions can be very precisely constructed due to the largeness of both ensembles;
\item[b.] The observed entropy of a piece of associative information would most likely be located at the mode of alternative entropy distribution, while the mode of the null entropy distribution would be located at the entropy value of the marginal distribution of ${\cal Y}$;
\item[c.] The overlapping area of alternative and null entropy distributions is the proper evaluation of the finite sample precision of this associative information piece, which is equal to the minimum sum of Type-I and Type-II errors. Based on this explicit finite sample precision, the meaning of signal strength becomes realistic and scientific;
\item[d.] While the so-called p-value is evaluated as the area under the null entropy distribution beyond the observed entropy toward the extreme direction away from the overlapping area direction. As illustrated in this generic setting, this p-value is most often too optimistic.
\end{description}
These four essential messages mark the critical differences between the SDA and all statistical analysis. The most important fabric of such differences is that scientific explanation is straight forward for any piece of associative relational information based on these four messages. We are confident that this is what has been argued and devoted by A. N. Kolmogorov in his 1983 paper via his randomness-proper concept in data analysis \cite{kolmogorov}.

In summary, a proper evaluation of the strength of signal of any piece of associative pattern information pertaining to a category of a covariate feature or feature-set can only evaluated by comparing with the empirical marginal distribution of ${\cal Y}$ offered in a contingency table, while such a comparison must be done by pertinently retaining the same sample sizes to the targeted observed covariate category. Only such comparison can render a proper reliability check that can bring out the proper, true scientific meaning of its signal strength. Hence, the capability of constructing a reliability checking protocol for any CEDA computed piece of associative information via a contingency table or equivalent histogram format pertaining to the category of targeted covariate feature or feature-set is tightly tied to the answer to a simply worded question: {\bf What and where is the relevant randomness in an observed contingency table?}

\subsection{Illustrations of robust associative information content with finite sample precisions in Iris .}
After addressing the previous question based on a contingency table pertaining to a covariate feature or feature-set, the next critical question is: How to extract associative relational information pertaining from a covariate feature or feature-set in full with least complexity? Via the Iris example, we address this seemingly complicate question starting from demonstrating somehow robust content of associative relational information with respect to various choices of HC-tree based contingency tables pertaining to one covariate feature or feature-set.

\begin{figure}[h!]
 \centering
\includegraphics[width=1.0\textwidth]{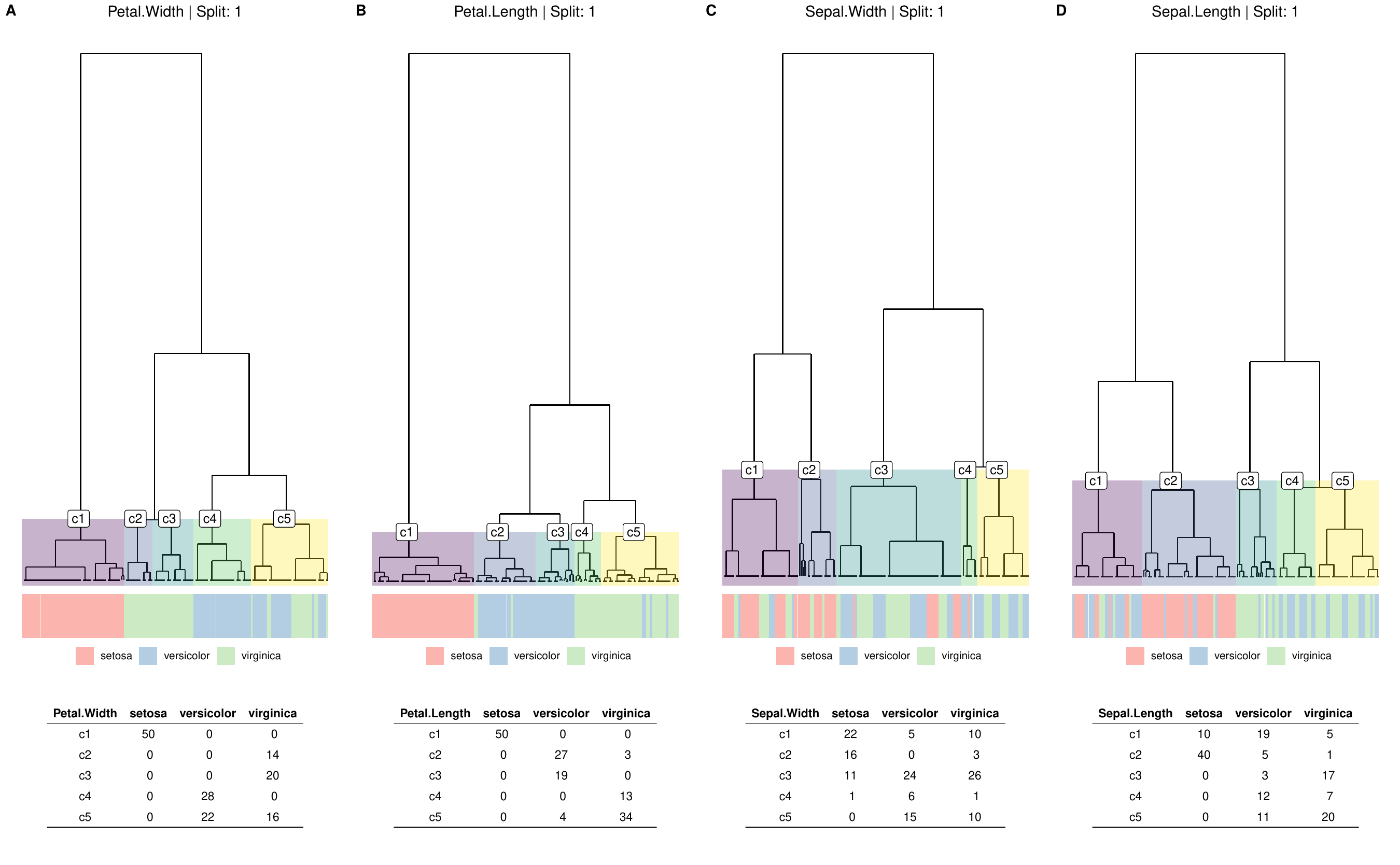}
\caption{Histogram-based contingency-tables of four features of Iris data. }
\label{Iristable}
 \end{figure}

In Fig~\ref{Iristable}, four HC-tree based contingency tables are respectively constructed for the four 1D features. Each HC-tree is uniformly cut at the 5 cluster tree-level. Thus, the five resultant contingency tables are in the $5\time 3$ format. Here the five categories of each $X_k$ are arranged along the row-axis, while the three species are arranged along the column-axis. By comparing this set of four contingency tables with the other set of four contingency table shown in Table~\ref{tab:Iristable0}, we can see to a great extent of robustness of associative relational information offered by anyone of the four 1D features. Across these four pairs of contingency tables, we can see that some column-vectors in Table~\ref{tab:Iristable0} with similar information contents are indeed merged into one row-vector in Fig~\ref{Iristable} that retains very much the same associative information content. This robust phenomenon is seen through the fact that these four HC-trees are the same in both figures. As would be described in detail after completing the illustration of finite sample precision below, this comparison indeed points to one protocol of dynamically exploring associative relational information through the tree branching process.

Take the $5\times 3$ contingency table $HCT[X_3;{\cal Y}]$ with $X_3$ (Petal length), as shown in the 2nd column-panel in Fig~\ref{Iristable}, as an example for purpose of explicitly and realistically illustrating Kolmogorov's randomness proper concept. With respect to $HCT[X_3;{\cal Y}]$, the alternative randomness underlying the observed contingency table $HCT[X_3;{\cal Y}]$ and the null randomness underlying simulated $5\times 3$ contingency tables, say $HCT[X^*_3;{\cal Y}]$, where the categorical variable $X^*_3$ has the same row-sum randomness, or the marginal categorical distribution as that of $X_3$, but is independent of ${\cal Y}$.

This alternative randomness would become visible when we operationally mimic the observed contingency table. The left column-panels of Fig~\ref{summary} displays the generative operations for this alternative randomness. That is, a large ensemble of mimicries of an observed contingency table should embrace potential signals so-called alternative randomness. While this null randomness indeed embrace all potential non-signals within a large ensemble of null contingency tables. The right column-panels of Fig~\ref{summary} displays the generative operations for this null randomness. An analytic view of this null randomness' role can be seen as follows.
\begin{eqnarray*}
E[n^*_{ij}]&=&150\times \frac{rows_i}{150} \times \frac{cols_j}{150},\\
&=&150\times Pr[X_3=C_i]\times Pr[{\cal Y}= ID[j]],\\
&=& 150\times Pr[(X_3, {\cal Y})=(C_i, ID[j])]
\end{eqnarray*}
where $rows_i$ is row-sum of row-$C_i$ and $cols_j$ is the column-sum of $ID[j]$-column. Recall here, the alternative and null randomness pertaining to the observed contingency table $HCT[X_3;{\cal Y}]$ has been summarized in Fig~\ref{summary}.

Based on the alternative and null randomness, we can respectively generate two ensembles of $5\times 3$ contingency tables of any size, say $B$. By picking one contingency table respectively from each of these two ensembles, we emphasize that the meaning of each row-specific associative pattern is brought out under the setting where the row-specific alternative and null randomness share the same expected sample size.

\begin{figure}[h!]
 \centering
\includegraphics[width=1.0\textwidth]{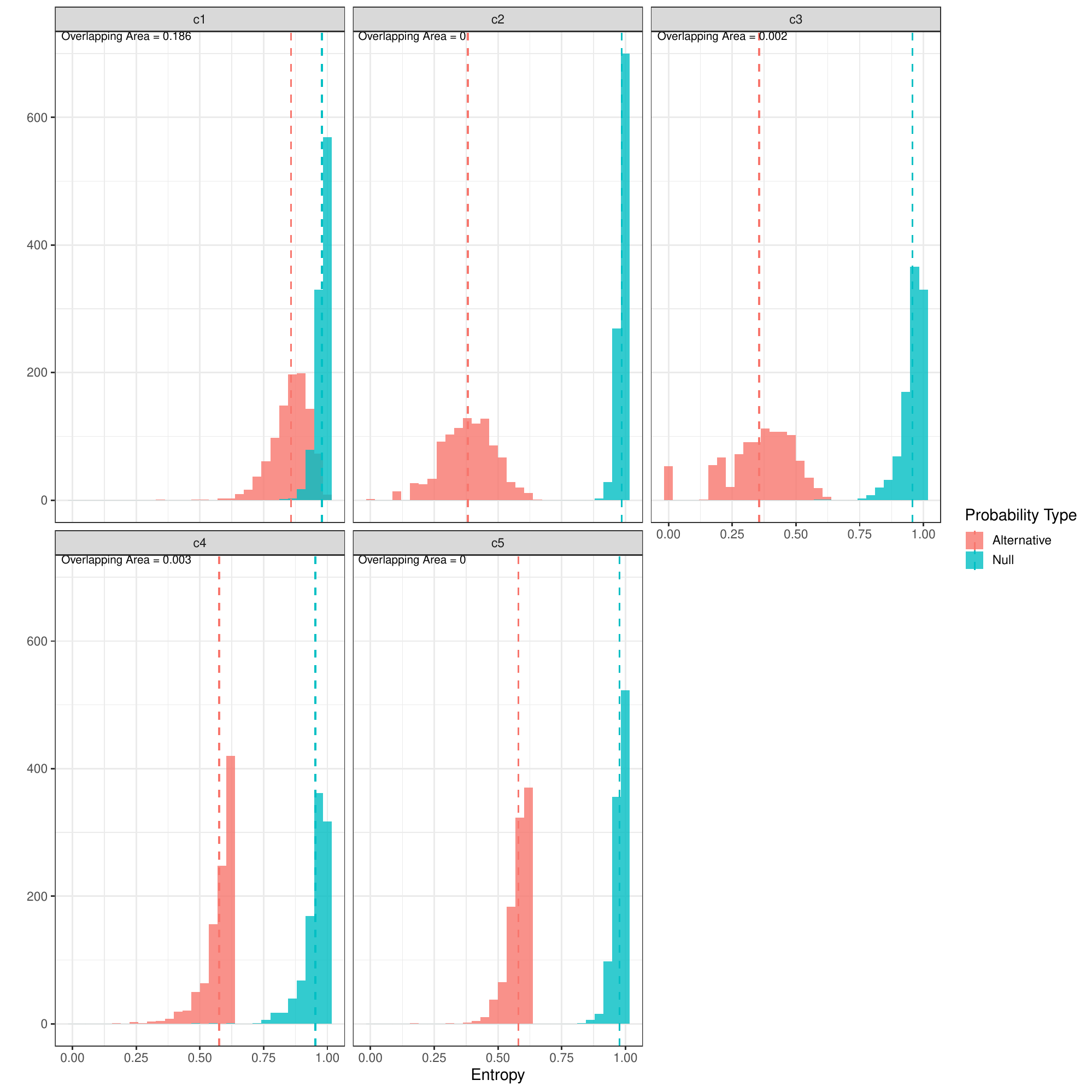}
 \caption{Illustrative overlapping areas of alternative (in red) and null (in blue) entropy histograms based on the row of five cluster: C1-C5 of Sepal Width. The five zero or nearly zero P-values are calculated at the observed row-wise entropy, which typically coincides with the mode of alternative entropy distribution.}
\label{overlapping}
 \end{figure}

Likewise, we can construct a pair of alternative-vs-null ensembles for any features or feature-sets. For instance, we easily switch to $X_2$ ( Sepal Width), and we have two collections of Shannon entropy with respect to each row-$C_i$ denoted as follows:
\begin{eqnarray*}
\{H^{(b)}[{\cal Y}|X_2=C_i]|b=1,..., B\}\\
\{H^{(b)}[{\cal Y}|X^*_2=C_i]|b=1,..., B\}.
\end{eqnarray*}
Based on these two collections of Shannon entropies, two histograms or densities can be built as illustrated in Fig~\ref{overlapping}. It is reiterated once more that the observed entropy is located at the mode of alternative entropy distribution, while the mode of the null entropy distribution is the entropy of marginal distribution of ${\cal Y}$ being equal to $\log 3$. Due to being unimodal for both entropy distribution, the overlapping area naturally is the minimum sum of type-I and type-II errors. This overlapping area is taken as the S-N ratio of the signal strength of associative relation of the $X_2=C_i$ to ${\cal Y}$. It is the finite sample precision that is visible, explicit and fully explainable.

According to Fig~\ref{Iristable}, all four 1D features ${\cal X}=\{X_1, X_2, X_3, X_4\}$ are categorized with 5 categories marked as $\{C_1, C_2,.., C_5\}$. The four corresponding contingency tables together provide 20 pieces of 1-feature-categories. It is obvious that all 50 irises of $setosa$ are characterized by the $C_1$ of Petal length ($X_3$) and $C_1$ of Petal width ($X_4$). Both are respectively the clusters of the smallest entropy measurements of both features. Further, the two Iris populations of $versicolor$ and $virginica$ are somehow separated into two branches of HC-trees of Petal length ($X_3$) and Petal width ($X_4$), respectively. Via HC-tree of Petal Length, the branch encoded as R1R2 consisting of clusters: $C_4$ and $C_5$, is mainly for $viginica$, the branch encoded as R1L2 consisting of clusters: $C_2$ and $C_3$, is nearly for $versicolor$ only. Similar degrees of separations are found via branching of HC-tree of Petal Width. Its $C_5$ consisting of 0 $setosa$, 22 $versicolor$ and 16 $virginica$ still would give rise to strong signal. With high signal strength, these ten 1-feature-categories of $X_3$ and $X_4$ are highly potential to pass their individual reliability checks. Such check would be based on their finite sample precisions via overlapping areas of their own idiosyncratic pair of alternative and null entropy distributions. Any such candidate passing the reliability check would be termed major 1-feature-categories, or major feature-category of order-1.

Collectively, we can display the above results based on the aforementioned ten 1-feature-categories via a so-called binary bipartite network. This binary bipartite network has a format of $150\times 10$ matrix lattice: all 150 Iris subjects are arranged along the row-axis, and the ten 1-feature-categories are arranged along the column-axis. The matrix's binary entries are all Iris-subjects' participating memberships of ten 1-feature-categories. When we permute row-axis and column-axis, we can expect to reveal the exclusive separation of 50 $setaso$ via two columns of $C_1$ of Petal width and length. This separating block in the matrix lattice would be confirmed blow. Various blocks can be likewise formed for members of $versicolor$ and $virginica$. We term this permuted bipartite matrix lattice a heatmap. See a simple illustration in the next subsection.

As the key platform for manifesting SDA results, this heatmap would be used to indicates that, to a great extent, different subpopulations of study-subjects in a data set are explicitly characterized by different formation of blocks framed by clusters of study-subjects and clusters of major major feature-categories of various orders. This is just one preliminary outline of how SDA works for Class-informatics.

\subsection{Lessening information content complexity.}
Up to this point, SDA results based on only the four 1D features, each of which has predetermined numbers of clusters as shown in both Table~\ref{tab:Iristable0} and Fig~\ref{Iristable}. In fact, the color-coding of tree-leaves under the two HC-trees of Petal width and Petal Length, respectively, in Fig~\ref{Iristable} clearly indicate the necessity of partitioning on the cluster $C_5$ of Petal Length and Petal width in order to achieve more exclusive separations of $versicolor$ and $virginica$. In other words, this necessity also implies that more exclusive major 1-feature-categories need to be explored and discovered by changing the scheme of categorizing quantitative features. The fixed number of categories determined priori indeed is not an effective categorizing scheme.

A data-driven categorizing algorithm is developed below. Since each HC-tree is binary. Therefore, each of its branch can be encoded by using binary coding scheme: Lk and Rk, for left and Right subbranches at the k-th internal node along any branching process. For instance, $C_4$ and $C_5$ on HC-trees of Petal width in Fig~\ref{Iristable} are coded as R1R2C3 and R1R2R3, respectively. All stoping-rules of branch-splitting are defined by the following protocol called Splitting-Stopping-Rule(SSR) algorithm.
\paragraph{Splitting-Stopping-rule(SSR) algorithm:}
\begin{itemize}
 \item[I][Stop-going-forward:] For Stop-rule-$\#k$ with $k=1, 2, 3,..$, when both two thresholds: 1)finite sample precision via overlapping area being less then $\alpha$; 2) signal strength via entropy being less than $\beta$, are achieved simultaneously.
\item[II][Information-Gain:] When a Stop-rule-$\#k$ with $k=1, 2, 3,..$ is not fulfilled at a $k$-th internal node along a branching process, the branch under exploring will be split and then checked for information gain in the sense of achieving stronger signal strength. This checking of whether the splitting operation indeed creates significant extra information improving upon its signal strength before splitting is limited to the concerned branch-locality. The computing of checking and confirming extra information is performed by representing the splitting operation via a contingency table with 2 rows against the response-categories. Each of these two rows are checked respectively to see whether the overlapping area is less than a threshold $\alpha$ and signal strength becomes stronger by having an entropy being less than $\beta$ simultaneously? If both subbranches fail to achieve extra information on top of their ``mother'' branch, then we need to decide whether to further explore once more or not. But if either one of subbranch does achieve extra information, and still not-yet-achieving the Stop-going-forward-rule, then further splitting is suggested unless the Sample-size-rule is achieved.
\item[III][Sample-size:] Any branch with a sample size less than a threshold will have no more splitting operations.
\end{itemize}
These two thresholds: $(\alpha,\beta)$, are man-made choices for decision-making regarding scientists and data analysts' human explorations on SDA final results. Different complex systems under different settings are likely required to have different degrees of finite sample precision and signal strength reflected through different thresholds.

Here this algorithm is applied on Petal length variable with a snapshot given in Table~\ref{tab:splitting}. We can clearly see that three zero-entropy major feature-categories are found on this seemingly very uninformative feature.

\begin{table}[t]
\centering
\caption{Split-Stop algorithm example for identifying 1D categories of Petal Width in the Iris data with ($\alpha$,$\beta$)=(0.1,0.25).}
\label{tab:splitting}
\fontsize{8pt}{9pt}\selectfont
\begin{adjustbox}{width=\textwidth}
\begin{tabular}{lrrrrrrrrl}
\toprule
Petal.Width & setosa & versicolor & virginica & Total & Entropy & EntropyOVA & furtherSplitIG & furtherSplitOVA & treeTraceBack \\
\midrule

\multicolumn{10}{l}{\textbf{Split \#1}} \\
\midrule
c1 & 50 & 0 & 0 & 50 & 0.000 & 0.000 & 0.000 & 1.000 & L1 \\
c2 & 0 & 50 & 50 & 100 & 0.631 & 0.000 & 0.298 & 0.000 & R1 \\

\midrule
\multicolumn{10}{l}{\textbf{Split \#2}} \\
\midrule
c1 & 50 & 0 & 0 & 50 & 0.000 & 0.000 & 0.000 & 1.000 & L1 \\
c2 & 0 & 0 & 34 & 34 & 0.000 & 0.000 & 0.000 & 1.000 & R1.L2 \\
c3 & 0 & 50 & 16 & 66 & 0.504 & 0.000 & 0.147 & 0.001 & R1.R2 \\

\midrule
\multicolumn{10}{l}{\textbf{Split \#3}} \\
\midrule
c1 & 50 & 0 & 0 & 50 & 0.000 & 0.000 & 0.000 & 1.000 & L1 \\
c2 & 0 & 0 & 34 & 34 & 0.000 & 0.000 & 0.000 & 1.000 & R1.L2 \\
c3 & 0 & 28 & 0 & 28 & 0.000 & 0.000 & 0.000 & 1.000 & R1.R2.L3 \\
c4 & 0 & 22 & 16 & 38 & 0.620 & 0.000 & 0.162 & 0.072 & R1.R2.R3 \\

\midrule
\multicolumn{10}{l}{\textbf{Split \#4}} \\
\midrule
c1 & 50 & 0 & 0 & 50 & 0.000 & 0.000 & 0.000 & 1.000 & L1 \\
c2 & 0 & 0 & 34 & 34 & 0.000 & 0.000 & 0.000 & 1.000 & R1.L2 \\
c3 & 0 & 28 & 0 & 28 & 0.000 & 0.000 & 0.000 & 1.000 & R1.R2.L3 \\
c4 & 0 & 17 & 3 & 20 & 0.385 & 0.001 & 0.002 & 0.930 & R1.R2.R3.L4 \\
c5 & 0 & 5 & 13 & 18 & 0.538 & 0.012 & 0.171 & 0.237 & R1.R2.R3.R4 \\

\bottomrule
\end{tabular}
\end{adjustbox}
\end{table}

This algorithm is applicable for 1D quantitative feature as well as a feature-set consisting any number of 1D quantitative feature. Since its basis is a HC-tree. Such a tree can be built by the Hierarchical Clustering algorithm applied on data set of any dimensionality.

\begin{figure}[h!]
 \centering
\includegraphics[width=1.0\textwidth]{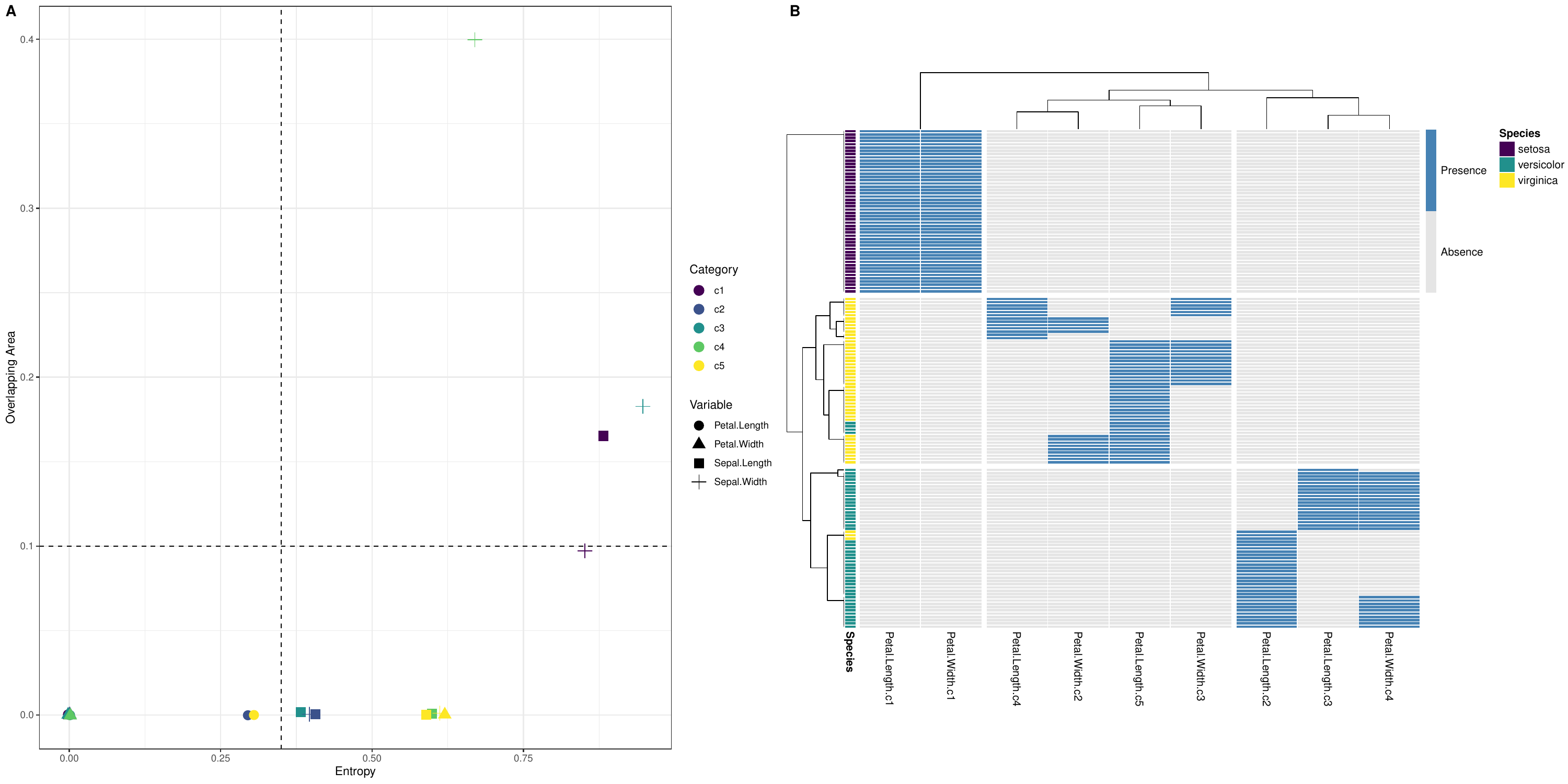}
 \caption{Plot of overlapping area against entropy of all 1D feature-categories (A) and heatmap of selected major 1-feature-categories (B). }
\label{sdafixed}
 \end{figure}

To illustrate Iris example with focus only on the four 1D features for expositional simplicity, the bivariate vectors: (observed entropy, overlapping area), are computed and reported in the panel (A) of Fig~\ref{sdafixed}. Here, the overlapping area is taken as the S-N ratio of any 1-feature-category, while the observed entropy of any feature-category is perceived as signal strength because of its degree of exclusiveness of memberships within a feature-category. That is, a zero entropy feature-category has all its memberships being exclusively belonging to one $ID[j]$ of  ${\cal Y}$. By choosing a threshold for overlapping area, say 0.1 and another one for the entropy, say 0.25, we select a set of so-called major 1-feature-categories, also called major feature-category of order-1. Further, we construct a binary heatmap of presence and absence of subject-memberships of each selected major feature-category. That is, as all subjects being arranged along its row-axis, while each column of this heatmap is 0 and 1 memberships across all subjects as shown in Panel (B) of of Fig~\ref{sdafixed}. This bipartite network construction clearly characterizes each individual subject with a binary vector of participation along the collection of selected major 1-feature-categories. Each subject's binary row-vector is termed as its heatmap specific individual character-landscape.

Here, a heatmap based on major feature-categories of order-1 is constructed for illustrative purpose only. For one Re-Co dynamics, we propose to build only one heatmap that collects and displays all major feature-categories of all orders, not just order-1, as would be shown in the next section. Such a heatmap will effectively reveal individual character-landscape for each study subject involved in a particular Re-Co dynamics. More specifically, such a heatmap's individual character-landscapes collectively constitute a topological neighborhood system upon the subject-space via simple Euclidean distance that defines the similarity among all these binary row vectors. As such this topology is a basis for information outlier detection. Further, such a topological neighborhood system indeed also provides a basis for a realty check upon the coupled response-ID annotations.

Equally important is that such a heatmap also provide a platform for visualizing various forms of mechanistic dependence among different feature-sets of different orders that essentially point to involving various mechanisms underlying this Re-Co dynamics. That is, such a heatmap becomes unique window for peeking into the underlying dynamics. These functional aspects of essential merits of SDA-based heatmap will constitute a basis for good explanations for the Taxonomy on Iris, as would be seen in the next section. Next we continue to illustrate the construction of heatmap containing major feature-categories of order-2 and higher under the Re-Co dynamics from ${\cal X}$ to ${\cal Y}$.

\subsection{High order interacting effects without curse of dimensionality} 
In this subsection, we elaborate one great advantage of SDA: the capability of exploring high order interacting effects. Such effects are primary mysteries in all sciences. Though the number of its functional forms is supposedly to be finite for any interacting effects of a fixed order in sciences, their diversity and mostly unknown forms make such effects rather mysterious. Such mysteries can't be handle correctly or scientifically by any top-down approaches, such as Statistical Modeling or Machine Learning (ML). Especially, the often used product-form of interacting effect of order-2 is ad hoc. Such ad hoc components make methodologies and techniques in both fields not only inadequate from the fundamental perspectives of Data Analysis, but also fictitious or even detrimental from the scientific perspectives of extracting data's authentic information content.

High order interacting effects can be succinctly described as: Whatever kinds of dependence pattern based major feature-categories of any orders under any focal Re-Co dynamics. Exploring such effects heavily relies on the capability of being able to correctly and precisely fuse any number of quantitative or categorical features, or mixed of both types, without being subject to the effects of Curse of Dimensionality. This is one vital capability thoroughly used throughout the entire developments of SDA. See such capability in CEDA of SDA in \cite{FCC23} and real-world applications in \cite{omotayo,michelon}.

First, under the quantitative setting of Iris data, SDA effectively explores dependence patterns among quantitative features simply by applying HC-algorithm. This widely used clustering algorithm is implemented with a choice of distance or dissimilarity measure and a modula. It is as simple as a function in R- and Python programming languages. That is, SDA employs HC-algorithm to fuse and then categorize a target set of multiple features into one 1D categorical feature through its resultant HC-tree. It is essential to reiterate that this HC-algorithm is a simple way of preserving and summarizing the dependence patterns with no limits regarding the dimensionality of any focal feature-set. And its HC-tree enables the Splitting-Stopping-rule(SSR) algorithm to freely explore potential subbranches having exclusive response-IDs. As such SDA is well-equipped to effectively explore high order interacting effects.

Nonetheless, when the dimensionality $K$ of ${\cal X}$ is large, the total number of all possible feature-sets is of the order of $2^K$, which can be astronomical. Which feature-sets are potential? Only if it is necessary for the sake of reducing computing cost, this issue can be resolved by building a network with pairwise-linkages, and then the focus is placed only on the collection of network-communities. Such a network construction is proposed to be based on mutual conditional entropy (MCE) due to existential heterogeneity among features \cite{CCF22a,CCF22b,CCF22c}. The MCE is calculated as follows:
\[
MCE[X_1, X_2]=\{\frac{H[X_1|X_2]}{H[X_1]}+\frac{H[X_2|X_1]}{H[X_2]}\}/2,
\]
where $\frac{H[X_1|X_2]}{H[X_1]}$ measures the proportion of the conditional entropy as remaining entropy the of $X_1$ given the information of $X_2$ relative to marginal entropy of $X_1$. MCE is related to the mutual information $I[X_1, X_2]=H[X_1]-H[X_1|X_2]=H[X_2]-H[X_2|X_1]$. However, without the proper re-scaling like what has been done in MCE, $I[X_1, X_2]$ is not a proper association measure. Since $I[X_1, X_2]\leq \min\{H[X_1], H[X_2]\}$.

By using $MCE[X_k, X_{k'}]$ as the linkage between two nodes $X_k$ and $X_{k'}$, this weighted network is built with $K$ nodes. Its multiscale community structures can be sought out, see \cite{CF12,FC14}. Each community signals a specific kind of dependence among members of the network community. The total number of communities, say $K_c$, upon such a network is likely much smaller than $K (>K_c)$. Hence the $2^{K_c} << 2^{K}$.  As such, each community found at a focal scale is a potential candidate feature-set for exploring high order interacting effects, so are the interacting effects among all communities of all possible orders. A Taiwanese Bankruptcy data set with 95 1D quantitative features is successfully analyzed in this fashion \cite{yidalin}.

As for this illustrative Iris data, SDA only faces 11 possible feature-sets of from order-2 to order-4. In SDA, pairwise, triplet-wise and quartet-wise feature-sets would go through fusing and categorizing computations via HC-algorithm and its HC-tree, and then their interacting effects of order-2 to order-4 are to be fully explored just like effects of the four 1D features discussed in the previous subsection.

In summary, theoretically speaking under the quantitative setting, this illustration of SDA on Iris data indeed can be easily extended to any quantitative feature-sets of any sizes, as shown in \cite{omotayo,michelon,yidalin}. Given that HC-algorithm and its HC-tree can be applied to fuse and categorize any quantitative feature-sets onto a 1D categorical variable, all SDA computations discussed in the previous subsection become applicable. Thus, SDA would not be hindered by the dimensionality of any quantitative feature-set. This is the way of studying and exploring high order interacting effects. The importance of such capability of peeking into high order interacting effects can not be underestimated in all sciences.

\subsection{Outlier detection}
In this subsection, we reiterate our emphasis that outlier detection indeed is the prerequisite for all forms of prediction. This prerequisite has been ignored completely partly due to the lack of systematic protocols in literature of Statistics and Machine Learning and A.I., and partly due to intentional as well as unintentional misunderstanding of the goal of any classification. In fact, its foundation, function and role of outlier detection naturally becomes self-evident from the taxonomy perspective. We systematically illustrate such a perspective all along the process of carrying out SDA in this subsection. This more than six decades old issue has not yet been solved so far to our best of knowledge. But we rigorously resolve it via SDA here.

Let a non-annotated targeted subject be denoted as $(x^@, y^@)$ with $y^@$ being unknown. Based on the observed data set of $({\cal X},{\cal Y})$, SDA's realistic and intuitive outlier detection task on $x^@$ would be outlined from two aspects: one geometric and one information. The geometric aspect is referring to whether $x^@$ could fit into all point-cloud-geometries of all vital, not necessary all, feature-subsets specific projections of ${\cal X}$, while the information aspect is whether $x^@$ could fit into the constructed heatmap via SDA under the spectrum of Re-Co dynamics. As recalled here, such a series of four heatmaps display a collective of computed and confirmed pieces of associative information of all orders under Re-Co dynamics: ${\cal X}$ to ${\cal Y}$ and to three ${\cal Y}_{sub}$.

For geometric outlier detection, a point-cloud-geometry of a feature-subset ${\cal X}_{sub}=\{X_{k_1}, X_{k_2},..., X_{k_h}\}$ is referring to its $h$-dim empirical distribution in a form of scatter plot in $R^h$ under a quantitative setting. And a vital feature-subset is referred to a feature-set that gives rise to at least one selected major feature-category listed as one column in a constructed heatmap. If ${\cal X}_{sub}$ is a vital feature-set, then the $h$-vector: $(x^@_{k_1}, x^@_{k_2},..., x^@_{k_h})$, must fit into the manifold of ${\cal X}_{sub}$ to be declared as non-geometric outlier. If $x^@$ is not an outlier with respect to all vital feature-sets involving in the heatmap, then its information outlier status is considered as follows. This $x^@$ will give rise to a binary row-vector as its individual character-landscape pertaining to the column-axis of the heatmap. If $x^@$'s individual character-landscape can fit into the topological system hosted on heatmap's row-axis by being accepted into a study-subject's neighborhood, then it is not an information outlier.

When performing both kinds of outlier detections, we make use of the same technical scheme. To check whether a targeted object-vector could fit into a targeted point-cloud-geometry or topological space, we starts with finding a small collection of candidate-neighbors within the topological space or geometry via a chosen distance measure. Usually Euclidean distance is used. With respect to the same distance measure, each candidate neighbor's neighborhood is also computed as a small collection of a fixed number of its nearest neighbors. This neighborhood would give rise to a range of distances. Then we check whether the distance between the targeted object-vector and this candidate neighbor is within the range or not.  This is how we determine whether this targeted object-vector could fit into this candidate neighbor's neighbor. If none of its candidate neighbors could include this targeted object-vector into their individual neighborhoods, then it is taken as an outlier with respect to this geometry or individual character-landscape topological space. See such a scheme used in an example of determining Pistachio ripening stages in \cite{omotayo}.

Specifically speaking, we say that $x^@$ passes a geometric outlier detection check with respect to a vital feature-subset ${\cal X}_{sub}$ specific point-cloud-geometry in the sense that $(x^@_{k_1}, x^@_{k_2},..., x^@_{k_h})$ can be included within one of its candidate neighbor's neighborhoods. Thus, $x^@$ is declared as a geometric outlier if it fails at anyone of outlier-detection-tasks pertaining to any vital feature-subsets selected with respect to any constructed heatmaps.

In a quantitative setting like Iris data, it is essential to note that, if $x^@$ is concluded as a non-geometric outlier throughout a series of vital feature-sets' point-cloud geometries pertaining to a heatmap, along this serial outlier detection tasks, $x^@$'s component-vectors corresponding to this series of vital feature-sets are accordingly categorized. That is, each of such a component-vector would be categorized by inheriting the category of its candidate neighbor that accept and accommodate it into its neighborhood. As such a heatmap specific individual character landscape is derived for $x^@$, with which the task of information outlier detection could be proceeded. We reiterate that the geometric outlier detection task is necessary for the task of categorizing $x^@$ under any quantitative setting.

Only when $x^@$ is concluded as a non-geometric outlier, then we would proceed to consider whether it is an information outlier or not. This $x^@$ is declared as an information outlier when its heatmap specific individual character-landscape can not fit into this heatmap's topology in the sense that it is excluded from neighborhoods of each every study-subject's heatmap specific individual character-landscapes. On the other hand, if $x^@$ is a non-information outlier under a Re-Co dynamics, then $x^@$'s heatmap specific individual character-landscape will reveal its potential annotation for $y^@$ coming from the involved topological neighborhood. This is how SDA resolves the outlier detection problem. On the other hand, this is the SDA's stand on predictive inference.

As for this illustrative Iris data, SDA would respectively build four heatmaps with respect to all four Re-Co dynamics from ${\cal X}$ to ${\cal Y}$ or one of three ${\cal Y}_{sub}$. Each heatmap accommodates major feature-categories of order-1 to order-4. As such, there are multiple point-cloud-geometries for geometric outlier detection tasks according to each of four heatmaps, while there are four heatmap specific information outlier detection tasks. This is one window to see the essential merit of Class-informatics. Also, this is how the outlier detection problem is universally resolved.

\section{Iris' Class-informatics}
\begin{figure}[h!]
 \centering
\includegraphics[width=1.0\textwidth]{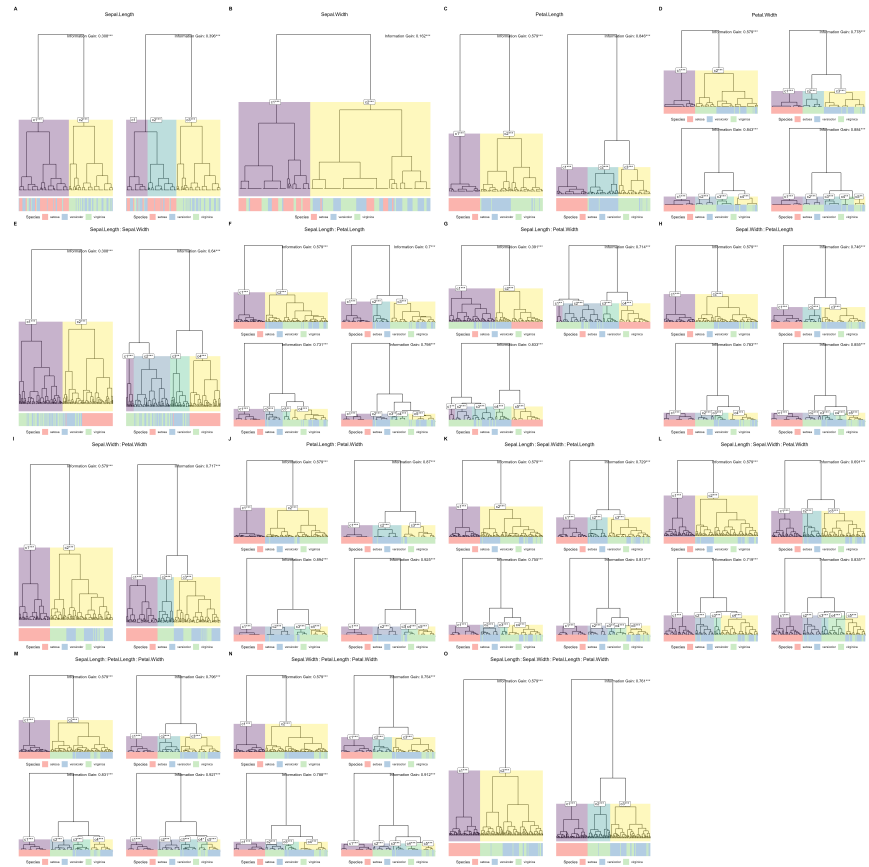}
 \caption{SDA computational results on $\{setosa, versicolor, virginica\}$ with all major feature-categories from order one to four. Each major feature-category's category-codes $C_k$ with $k=1,2...$ via HC-trees are explicitly listed. All columns of Heatmap in Fig~\ref{3iris} are defined.}
\label{SDAtree3iris}
 \end{figure}

\begin{figure}[h!]
 \centering
\includegraphics[width=1.0\textwidth]{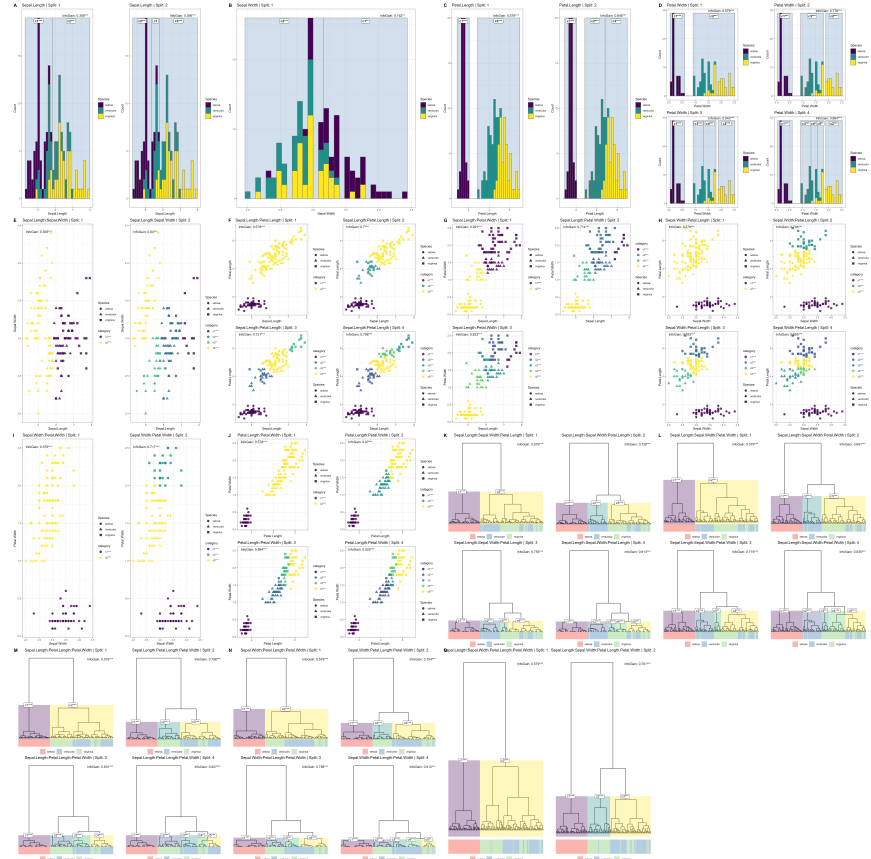}
 \caption{SDA computational results on $\{setosa, versicolor, virginica\}$ with all major feature-categories from order one to four. Each major feature-category's precise geometric definitions and biological meanings of category-codes $ck$ with $k=1,2...$ are explicitly listed. All columns of Heatmap in Fig~\ref{3iris} are explicitly explained.}
\label{SDAplot3iris}
 \end{figure}

Before explaining SDA resultant heatmaps and making corresponding biological conjectures, its is beneficial to take a second look at all panels in Fig~\ref{pairplot}. Along the four diagonal panels, the four $setosa$-specific density functions of the four features give rise to a very distinctive pattern: The spreads of the two $setosa$-specific density functions pertaining to Sepal length and width are evidently spreading wider than the spreads of the two $setosa$-specific density functions pertaining to Petal length and width. This pattern strongly indicates that measurements of Petal length and width of $setosa$-species are much more contracted than the measurements of Sepal length and width within the same Iris-species. This is the biological differences of order-1 effects attributed to Septal length and width and Petal length and width. This simple observation indeed is consequential in SDA computational results as shown in Fig~\ref{SDAtree3iris} and Fig~\ref{SDAplot3iris}.

In contrast, the biological differences of order-2 effects attributed to $\binom{4}{2}$ pairs of the four features: Septal length and width and Petal length and width, are seen through the six panels of 2D $setosa$-specific clouds or scatter-plots below the diagonal panels in Fig~\ref{pairplot}. The $setosa$-specific cloud pertaining to the bivariate (Sepal length, Sepal width) in the $2\times 1$ panel is much wider spreading than the $setosa$-specific clouds within the rest of 5 panels. Further, the $setosa$-specific cloud in the $2\times 1$ panel is much closer to the $versicolor+virginca$-specific cloud than that in the rest of five panels. Especially, the degree of contraction on 2D $setosa$-specific cloud is particularly striking as shown in the $4\times 3$ panel pertaining to bivariate (Petal length, Petal width). These simple observation would further make up vital mechanistic consequences in SDA computational results as also shown in Fig~\ref{SDAtree3iris} and Fig~\ref{SDAplot3iris}.

In summary, the $setosa$-specific 1D and 2D clouds or scatter-plots are clearly contracted to a much greater extent than the other two species' when involving with either Petal length or width. That is intuitive to say that both Petal length and width have dominant contracting effects operating within $setosa$ species. With these ideas in mind, the heatmap-based interpretations given in this and next sections would become biologically more evident, and the conjectures motivated by the contracting patterns would become meaningful and even realistic.

\begin{figure}[h!]
 \centering
\includegraphics[width=1.0\textwidth]{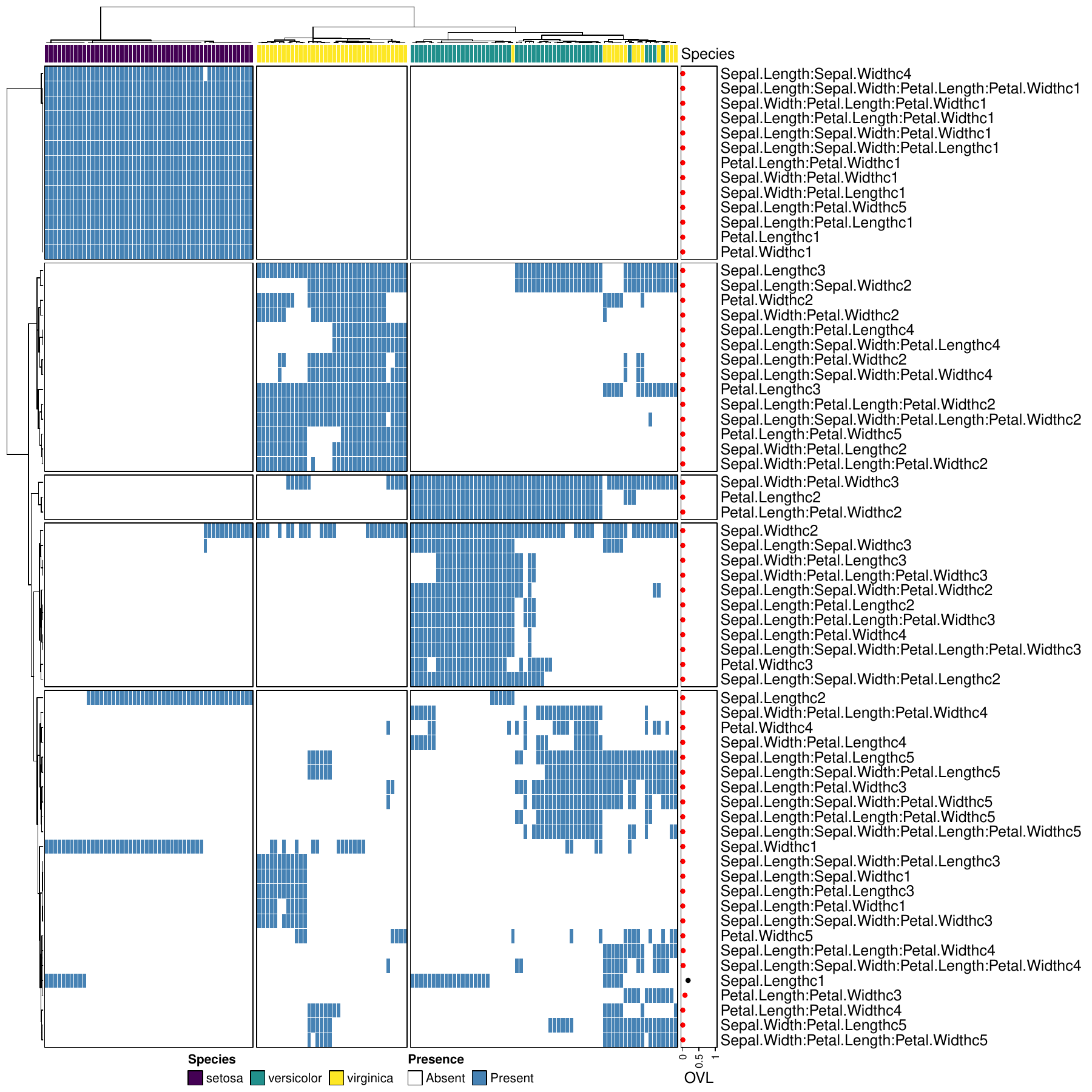}
 \caption{Heatmap of classification on $\{setosa, versicolor, virginica\}$ with all major feature-categories from order one to four. The precise geometric definitions and biological meanings of category-codes $ck$ with $k=1,2...$ can be found in Supporting Information(SI).}
\label{3iris}
 \end{figure}

We begin with constructing of the first part of Class-informatics under ${\cal Y}=\{setosa, versicolor, virginica\}$ setting. Within the Re-Co dynamics: from ${\cal X}$ to ${\cal Y}$, all feature-categories resulted from applying the Splitting-Stopping-rule(SSR) algorithm are defined and reported via HC-trees of all orders in Fig~\ref{SDAtree3iris} and via scatter plots of 1D and 2D in Fig~\ref{SDAplot3iris}. While Iris' binary memberships of all confirmed major feature-categories are collectively reported through a heatmap in Fig~\ref{3iris}. Along its row-axis of such a heatmap, all computed and confirmed major feature-categories of order-1 to order-4 are displayed, while all 150 Irises are arranged along the column-axis. Each row is the recording binary presence or absence of Iris memberships with respect to the one major feature-category. In other words, each column is a binary participation of one Iris across all selected major feature-categories, which is this Iris' individual character-landscape. Such individual Iris' characteristics is the basis for understanding the Iris complex system under study. Since such a heatmap is an informative bipartite network that is capable of bringing out interacting relationships embraced between Iris-subject space and major feature-category space.

All major feature-categories are selected with respect to a threshold for overlapping area being set at $0.01$ are marked with red dots, otherwise blue dots along the 151th extra column. This heatmap is framed by two HC-trees on both row- and column-axes resulted from an algorithm called Data Mechanics \cite{CF12,FC14}. Data Mechanics is an algorithm of applying HC-computing alternatively between column-axis and row-axis of heatmap's matrix-lattice. Such applications of HC algorithm employ modified distance measures of row- and column-vectors by adapting tree-structures found on column- and row-axes of previous run. For better recognitions and visualizations, a $5\times 3$ block-array is marked and displayed. Certainly more refined block-array can be displayed to reveal even finer patterns. What are readable, visible and explainable patterns in this heatmap via this $5\times 5$ block-array?

Before diving into the pattern information, we encode all blocks by employing the same branch-encoding scheme used in Fig~\ref{Iristable}. All interpretations and explanations pertaining to these blocks are referred to Fig~\ref{SDAtree3iris} and Fig~\ref{SDAplot3iris}. The five marked branches or cluster of major feature-categories on row-axis from top(T)-to-bottom(B)are:$\{T1, B1T2, B1B2T3, B1B2B3T4, B1B2B3B4\}$. Likewise the three branches or clusters of Iris on column-axis from the left(L)-to-right(R) are encoded as: $\{L1, R1L2, R1R2\}$. Each branch on either row or column axes is seen being characterized by a horizontal or vertical series of blocks.

For instance, the L1 cluster of Iris consisting of all 50 $setosa$ is characterized by 5 blocks encoded as: $\{L1\times T1, L1\times B1T2, L1\times B1B2T3, L1\times B1B2B3T4, L1\times B1B2B3B4\}$. The first block on this vertical series is an almost $50 \times 13$ array of 1's, while the rest of four block are almost all 0's. It means that all 50 $setosa$ Irises are present on almost all 13 major feature-categories, while absent in all major feature-categories belonging to the four branches: $\{B1T2, B1B2T3, B1B2B3T4, B1B2B3B4\}$. With all cluster-IDs shown in Fig~\ref{SDAtree3iris} and Fig~\ref{SDAplot3iris}, this is an aspect of explicit explanation of $setosa$ under the setting of ${\cal Y}=\{setosa, versicolor, virginica\}$. Detailed interpretations are given as follows. It is reiterated that all categories resulted from applications of Splitting-Stopping-rule(SSR) algorithm on of feature or feature-set marks are encoded by $C_k$ with $k=1,2...$ and their precise geometric definitions and biological meanings can be found in Fig~\ref{SDAtree3iris} and Fig~\ref{SDAplot3iris}.

\begin{description}
\item[L1-1.]$setosa$ can be seen from the characteristics of $T1$ branch, which consists of 13 major feature-categories of order-1 to order-4. All 13 major feature-categories (with red marks) are equipped with entirely and completely exclusive presences for $setosa$ and exclusive absences for $versicolor$ and $virginica$.
\item[L1-2.] Especially, the fact that two order-1 major feature-categories: Petal.Width.c1 and Petal.Length.c1, share entirely and completely exclusive presences for $setosa$ and exclusive absences for $versicolor$ and $virginica$ strongly indicate their dominant effects in the mechanistic dependence underlying biological dynamics of $setosa$.
\item[L1-3.] This dominance is fully and explicitly manifested through the fact that the presences of 50 $setosa$ and absences of 50 $versicolor$ and 50 $virginica$ are commonly shared exclusively by all 12 major feature-categories, all possible interacting effects involving either Petal.Width or Petal.Length: 5 of order-2, 4 of order-3 and one of order-4.
\item[L1-4.] The pattern information of item[L1-1] to item[L1-3] clearly indicates that $setosa$ as a species uniquely embraces a rather unique and rigid phrase of underlying dynamics pertaining to Iris in the sense of being locked in smallness of Petal.Width and Petal.Length.
\item[L1-5.] On the other hand, this defining phase of $setosa$ reveals almost zero degree of plasticity seen through the order-2 interacting effect of Sepal.Width:Sepal.Length.c4, which is a category indicating simultaneous largeness on both Sepal.Width and Sepal.Length.
\item[L1-6.] Regarding the rest of four blocks on this vertical series, the two blocks: $\{L1\times B1T2, L1\times B1B2T3\}$ are complete 0's blocks, while the rest of two clusters: $\{L1\times B1B2B3T4, L1\times B1B2B3B4\}$ would become completely 0's when take off four rows corresponding to 1-feature-categories of either Sepal.Width or Sepal.Length. As such the rigidity of biological phase pertaining to $setosa$ is clearly revealed.
\end{description}

After explaining the rigidity of biological phase pertaining to $setosa$, we turn to the two horizontal block series for the two branches on row-axis: $\{R1L2, R1R2\}$. The cluster $R1L2$ consists of $virginica$ exclusively, while $R1R2$ embraces entire $versicolor$ and a group of $virginica$  Irises. In fact, if we take a closer look, then it is evident that $R1R2$ could be further split into three sub-clusters:$\{R1R2L3, R1R2R3L4, R1R2R3R4\}$. The two sub-cluster $\{R1R2L3, R1R2R3L4\}$ consist of almost exclusive $versicolor$ Irises, while the sub-cluster $R1R2R3R4$ is a mixed of $versicolor$ and $virginica$. This fact together with exclusiveness of $L1$ and $R1L2$ strongly indicate the Taxonomic Hierarchy of these three Iris species.

Next, based on the heatmap in Fig~\ref{3iris}, we describe the characterizations of $R1L2$, which consists of slightly more than half of $virginica$ Irises, via the vertical chain of blocks: $\{R1L2\times T1, R1L2\times B1T2, R1L2\times B1B2T3, R1L2\times B1B2B3T4, R1L2\times B1B2B3B4\}$, framed by it and 5 clusters of major feature-categories as follows. Again, see Fig~\ref{SDAtree3iris} and Fig~\ref{SDAplot3iris} for precise geometric definitions and biological meanings of all involved major feature-categories.
\begin{description}
\item[R1L2-1.] The entirely zero $R1L2\times T1$ block is a complete opposite of the block $L1\times T1$. All its Irises are completely absent across the 13 major feature-categories of order-1 to order-4 belonging to the cluster $T1$.
\item[R1L2-2.] The majority of entries of $R1L2\times B1T2$ block are 1s indicating that its Irises are presence among 14 major feature-categories, which include all possible feature-combinations of all orders, except the order-1 Sepal.Length.
\item[R1L2-3.] The $R1L2\times B1B2T3$ block is nearly entirely 0s. The cluster $B1B2T3$ contains only three major feature-categories.
\item[R1L2-4.] The $R1L2\times B1B2B3T4$ block is completely 0s, except its 1st row: Sepal.Width.c2 indicating small measurements of it.
\item[B1T2-5.] The majority of $R1L2\times B1B2B3B4$ block are 0s. It contains a small block 1s. If the cluster $R1L2$ indeed is split into $R1L2L3$ and $R1L2R3$. This small block can be identified as being framed by $R1L2L3$ and a cluster of 5 major feature-categories of order-2 and order-3. They all involve Sepal.Length. This fact reveals the heterogeneity within $R1L2$ cluster of $virginica$. That is, the cluster $R1L2$ indeed can be split into $R1L2L3$ and $R1L2R3$ to better reveal its heterogeneity.
\end{description}

Next, we discuss the three sub-clusters of $R1R2$:$\{R1R2L3, R1R2R3L4, R1R2R3R4\}$, respectively. By being a cluster of nearly exclusive $versicolor$, the five blocks framed by $R1R2L3$ and five clusters :$\{T1, B1T2, B1B2T3, B1B2B3T4, B1B2B3B4\}$ on row-axis are sharply contrasting with the five blocks characterizing $L1$ and $R1L2$. The 1's blocks of $R1R2L3$ are against 0's blocks of $L1$ and $R1L2$, and vise versa. While the sub-clusters $R1R2R3L4$ and $R1R2R3R4$ retain the complementarily and distinctly contrasting block-characterizations. It is important to note that sub-cluster $R1R2R3R4$ is an overlapping part of $versicolor$ and $virginica$. This observation says that more information is needed to better separate these two species of Iris.

In summary, the block-structural patterns revealed upon this heatmap in Fig~\ref{3iris} manifest diverse component-wise mechanical dependence of multiscale nature. Here, each block: either presence or absence, stands for a scale-specific dependence across all its involving major feature-categories. Such mechanical dependence is visible by sharing the same cluster of Irises on the row-axis. Further, each Iris cluster on the column-axis is characterized by a horizontal series of blocks.  So, each Iris cluster is characterized by a series of mechanical dependence in a collective form of Irises' individual character-landscapes. The above mentioned patterns and characterizations are parts of Class-informatics of Iris data.

Since the heatmap's block-patterns are framed by both HC-trees on column- and row-axes. The multiscale nature of mechanical dependence becomes visible. Such multiscale characteristics are evidently seen through multiple clusters of $versicolor$ and $virginica$ Irises. Their patterns of mechanical dependence are visible, not too complex, but can't be narrated with simple expositions. Such a cluster can only be characterized by a vertical series of fine scaled blocks. As these block-based mechanical dependence has also more details of finer scale, each Iris' individual character-landscape is more complex.

It is not hard to imagine that each Iris' individual landscape would be much more complex if there were much more than three species of Iris involved in this data set. Specifically speaking, such complexity seen within this heatmap has a lot to do the fact that there are three signal-to-noise (S-N) ratios with significantly different sizes being simultaneously involving in this classification of $Y=\{setosa, versicolor, virginica\}$. The two S-N ratios of $Y_{sub}=\{setosa, versicolor\}$ and $Y_{sub}=\{setosa, virginica\}$ are much larger than the S-N ratio of classification of $Y_{sub}=\{versicolor, virginica\}$. This linkage between the visible complexity in heatmap and the heterogeneity of S-N ratios points out two versions of reality. First, the taxonomic task on analyzing the Iris data is clearly resolved.  Secondly, to make good use of such phenomenal S-N ratio differences, it is necessary apply SDA onto the three $Y_{sub}$ pairwise classifications on top of $Y=\{setosa, versicolor, virginica\}$ to further strengthen the content of Class-informatics of Iris data. Then, this necessity immediately faces the critical issue: How to effectively perform SDA on $Y_{sub}$?

\section{Iris' SDA systemic Class-informatics via ${\cal Y}_{sub}$ }
Here, via three pairwise comparisons of $Y_{sub}$ with heterogeneous S-N ratios, we would computationally explore associative patterns to demonstrate Class-informatics via mechanical dependence through patterns embedded within many facets of geometric manifolds among three species of Iris. That is, we expect that such geometry-based patterns coupled with mechanical dependence formations would enhance Class-informatics and sharpen our understanding on Iris taxonomy. It is somehow surprising that SDA on all three $Y_{sub}$ would, and must, respectively involve data of the nonparticipating species. In this fashion the three heatmaps: associative relational information pieces and geometric outlier statuses, would complete Class-informatics for Iris taxonomy.

\subsection{Class-informatics via ${\cal Y}_{1}$}
\begin{figure}[h!]
 \centering
\includegraphics[width=0.8\textwidth]{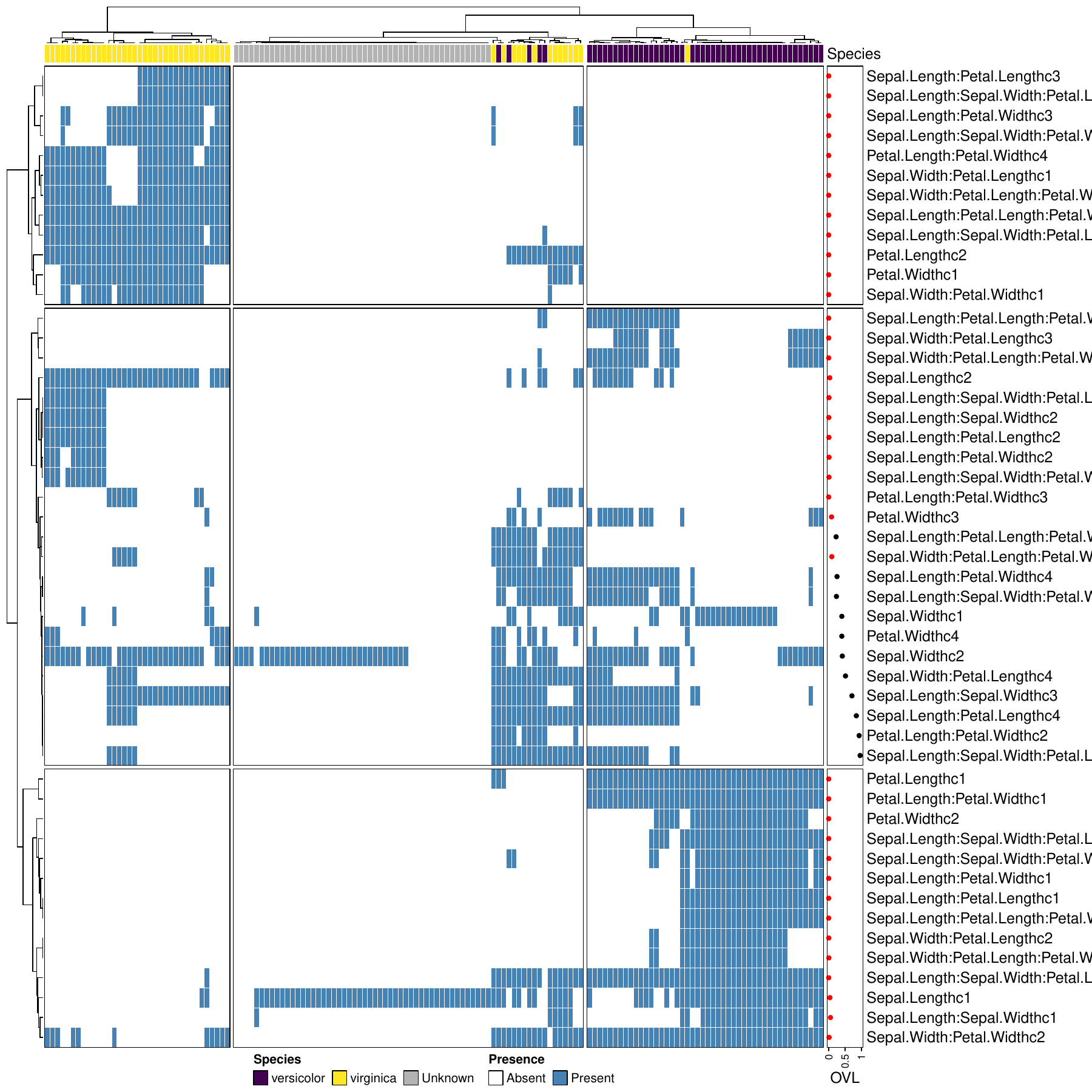}
 \caption{heatmap of classification on $\{versicolor, virginica\}$ all major feature-categories from order-1 to order-4 coupled with $setosa$'s information outlier status. Setosa Irises are marked as ``unknown''.}
\label{veviTose}
 \end{figure}

\begin{figure}[h!]
 \centering
\includegraphics[width=1.0\textwidth]{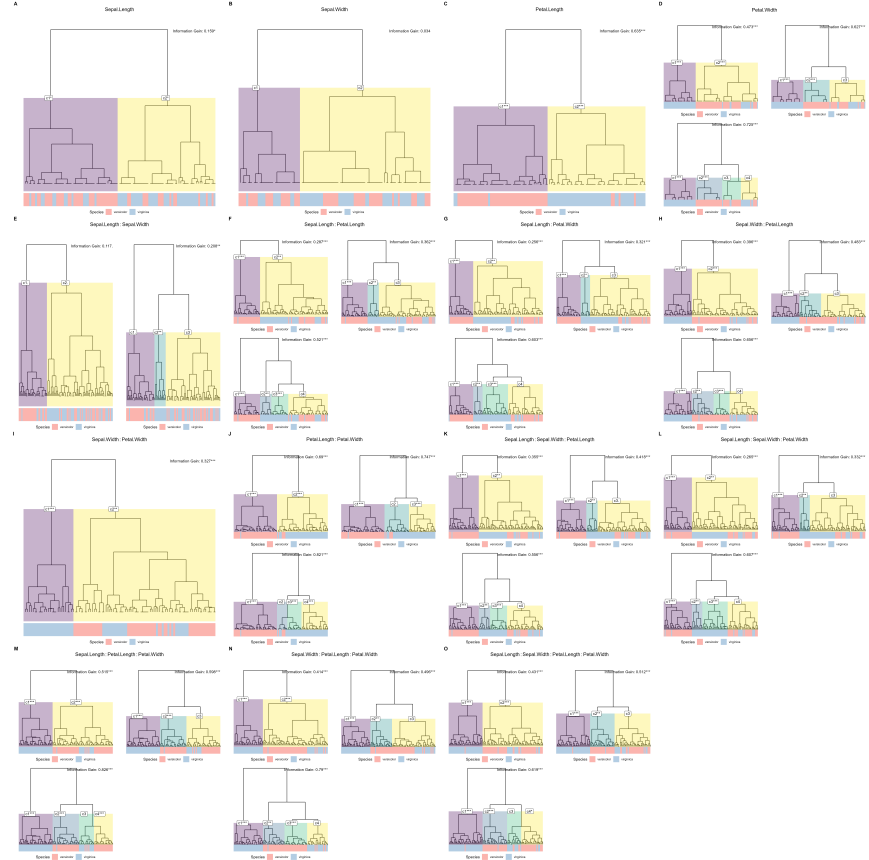}
 \caption{SDA computational results on $\{versicolor, virginica\}$ with all major feature-categories from order one to four. Each major feature-category's category-codes $ck$ with $k=1,2...$ via HC-trees are explicitly listed. All columns of Heatmap in Fig~\ref{veviTose} are defined.}
\label{SDAtreeveviTose}
 \end{figure}

\begin{figure}[h!]
 \centering
\includegraphics[width=1.0\textwidth]{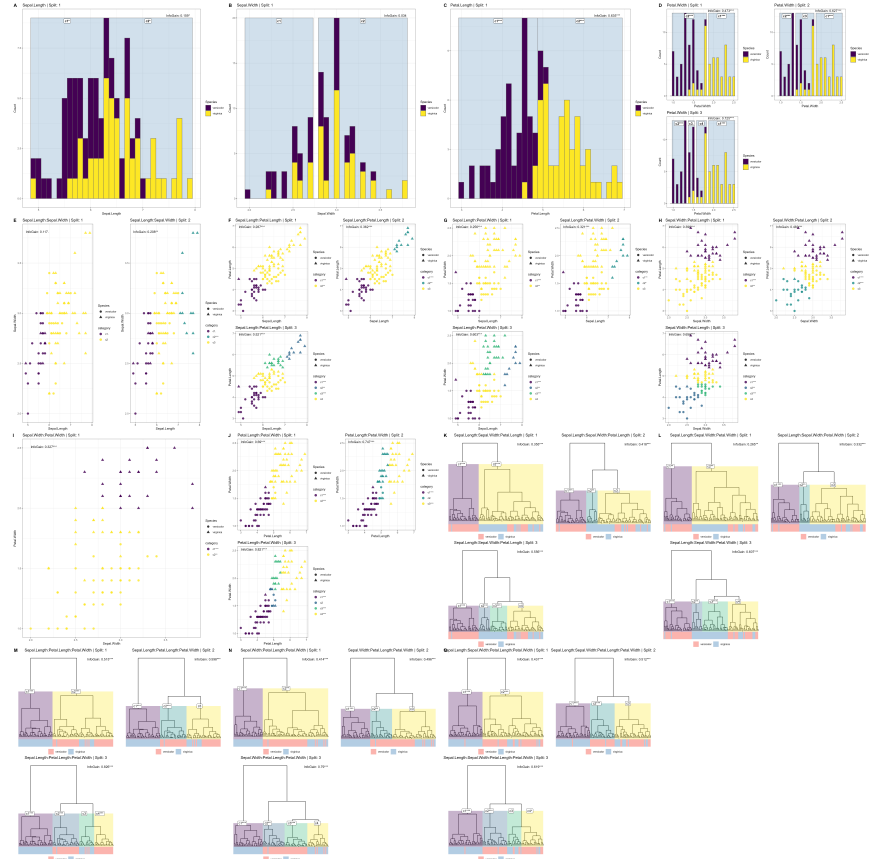}
 \caption{SDA computational results on $\{versicolor, virginica\}$ with all major feature-categories from order one to four. Each major feature-category's precise geometric definitions and biological meanings of category-codes $ck$ with $k=1,2...$ are explicitly listed. All columns of Heatmap in Fig~\ref{veviTose} are explicitly explained.}
\label{SDAplotveviTose}
 \end{figure}

We start subsystem-explorations via the Re-Co dynamics: from ${\cal X}$ to ${\cal Y}_{1}$, with ${\cal Y}_{1}=\{Ve, Vi\}$. The SDA computations for associative relational major feature-categories under this Re-Co dynamics are performed involving only 100 Iris-data from both species. That is, the 50 $setosa$ Irises are left out. Within the Re-Co dynamics: from ${\cal X}$ to ${\cal Y}_{1}$, all feature-categories resulted from applying the Splitting-Stopping-rule(SSR) algorithm are explored and reported via HC-trees of all orders in Fig~\ref{SDAtreeveviTose} and via 1D and 2D scatter plots in Fig~\ref{SDAplotveviTose}. While Iris' binary memberships of all confirmed major feature-categories are collectively reported through a heatmap in Fig~\ref{veviTose}.

At the first glance of the heatmap of Fig~\ref{veviTose}, the characteristic differences between $vesicolor$ and $viginica$ of global scale are slightly distinct from the differences seen in Fig~\ref{3iris}. The heterogeneity pertaining to $vesicolor$ and $viginica$ are respectively even more visible, and the overlapping between these two species are evident as well. Again, such species-to-species differences rest on computed major feature-categories of order-1 to order-4, which are bounded by various versions of mechanical dependence. The collective information outlier status of $setosa$-class within this heatmap is simply and completely exclusive: ``Absence from the all major feature-categories involved with Petal.Length or Petal.Width''. Class-informatics via mechanical dependence under ${\cal Y}_1$ are explained as follows. And all categories resulted from applications of Splitting-Stopping-rule(SSR) algorithm on of feature or feature-set marks are and their precise geometric definitions and biological meanings can be found in Fig~\ref{SDAtreeveviTose} and Fig~\ref{SDAplotveviTose}.

The heatmap of Fig~\ref{veviTose} is a bipartite network having all 150, not just 100, Iris arranged along the column-axis. Irises of $vesicolor$ and $viginica$ are color-coded, while $setosa$ Irises are marked as ``unknown with gray color-code''. Only the presence and absence of $vesicolor$ and $viginica$ memberships are involved in all selected and confirmed major feature-categories of order-1 to order-4, while the 50 $setosa$ Irises' information outlier statuses are included and arranged corresponding to all computed and confirmed major feature-categories arranged along the row-axis. These 150 binary vectors encoded with two different kinds of information contents are placed into one heatmap because they share the same row-axis. The essential merit of display is that this heatmap platform will allow us to clearly see the true meaning of information outlier. This merit is seen especially when the Data Mechanics algorithm is applied. The resultant two HC-trees respectively will frame both axes into readable, visitable and explainable block-structures.

The three major branches of HC-tree on column-axis are: $L1$, $R1L2$ and $R1R2$. The $L1$ and $R1R2$ are nearly exclusive for $viginica$ and $vesicolor$, respectively. The $R1L2$ should be split into: $R1L2L3L4$, $R1L2L3R4$ and $R1L2R3$. The $R1L2L3L4$ consists of 50 Setosa Irises characterized with their outlier statuses, while $R1L2L3R4$ and $R1L2R3$ are mixed of $viginica$ and $vesicolor$ with distinct characterizations.

\subsection{Class-informatics via ${\cal Y}_{3}$.}
\begin{figure}[h!]
 \centering
\includegraphics[width=0.8\textwidth]{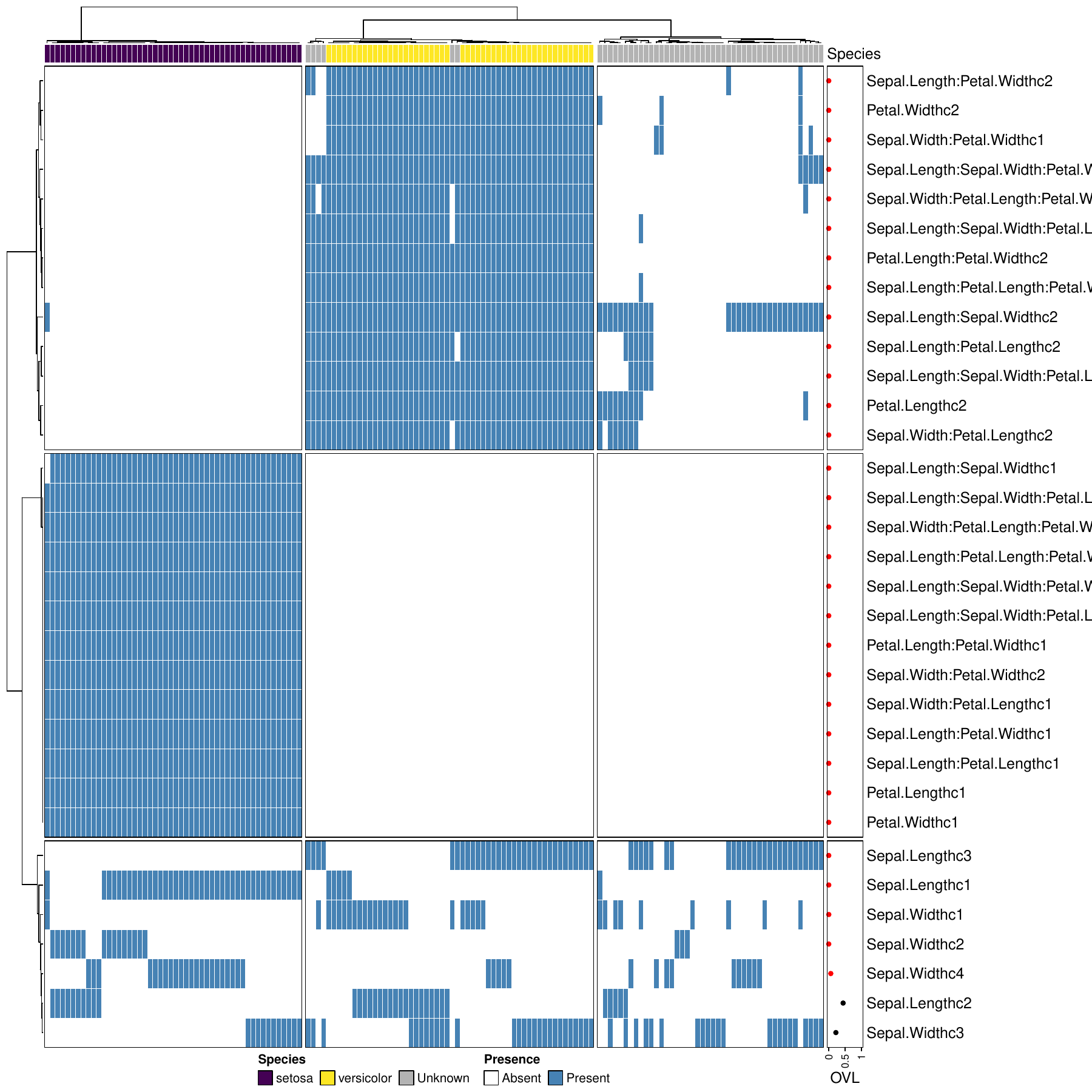}
 \caption{heatmap of classification on $\{setosa, versicolor\}$ all major feature-categories from order one to four coupled with $virginica$'s information outlier status. $virginica$ Irises are marked as ``unknown''.}
\label{seveTovi}
 \end{figure}

\begin{figure}[h!]
 \centering
\includegraphics[width=1.0\textwidth]{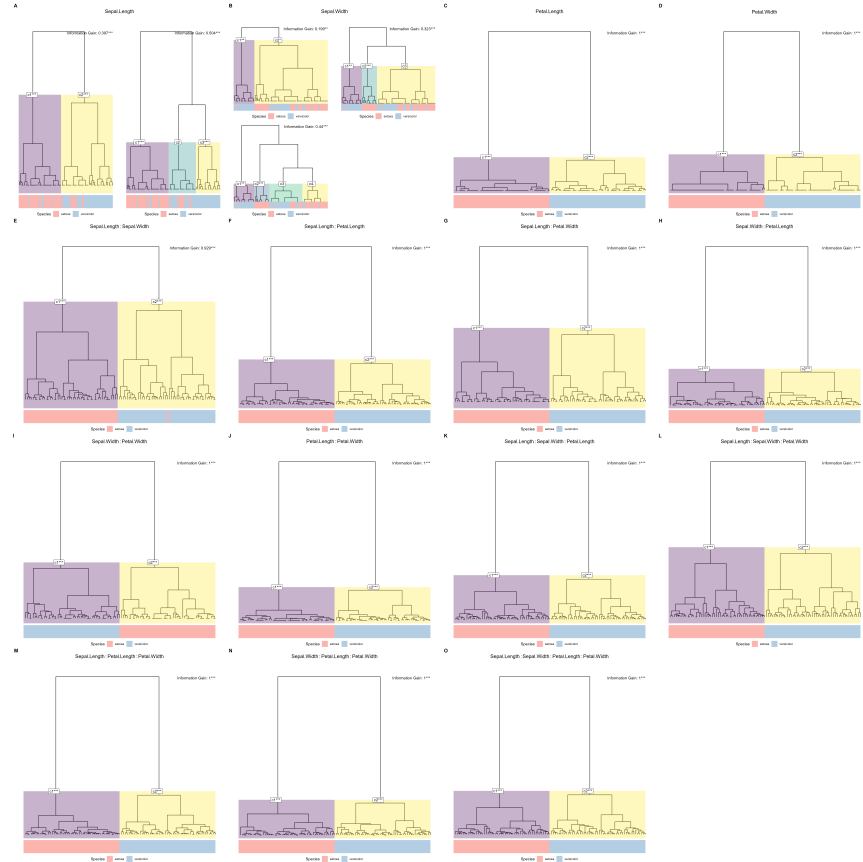}
 \caption{SDA computational results on $\{setosa, versicolor\}$ with all major feature-categories from order one to four. Each major feature-category's category-codes $ck$ with $k=1,2...$ via HC-trees are explicitly listed. All columns of Heatmap in Fig~\ref{seveTovi} are defined.}
\label{SDAtreeseveTovi}
 \end{figure}

\begin{figure}[h!]
 \centering
\includegraphics[width=1.0\textwidth]{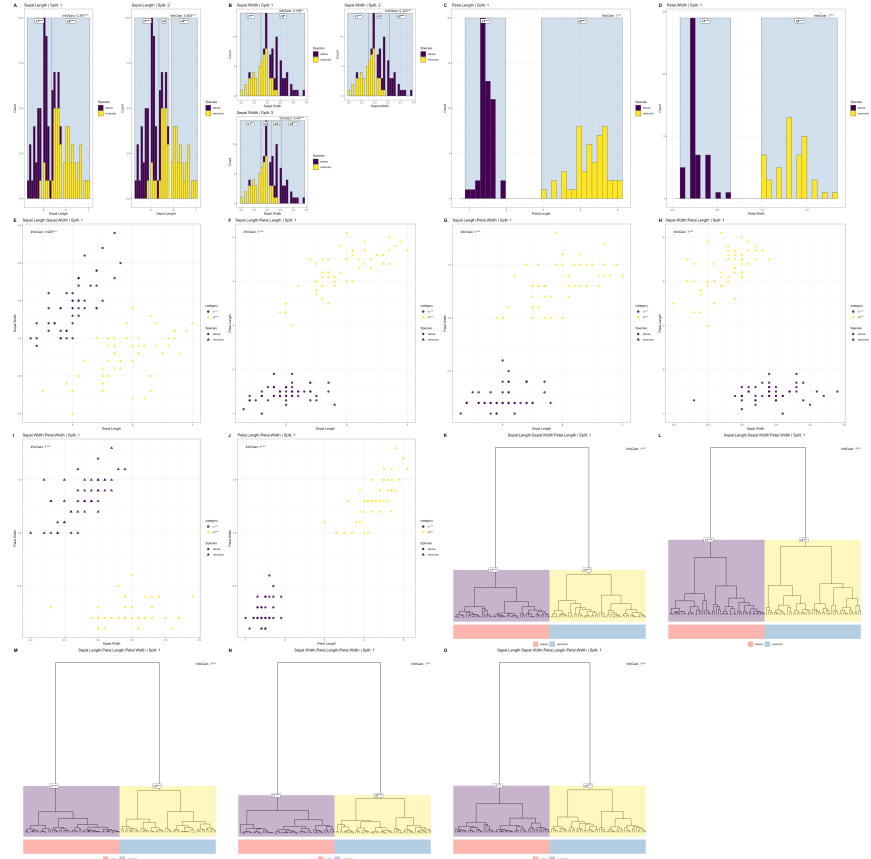}
 \caption{SDA computational results on $\{setosa, versicolor\}$ with all major feature-categories from order one to four. Each major feature-category's precise geometric definitions and biological meanings of category-codes $ck$ with $k=1,2...$ are explicitly listed. All columns of Heatmap in Fig~\ref{seveTovi} are explicitly explained.}
\label{SDAplotseveTovi}
 \end{figure}

Next we turn to subsystem-explorations via the Re-Co dynamics: from ${\cal X}$ to ${\cal Y}_{3}$, with ${\cal Y}_{3}=\{Se, Ve\}$. Within the Re-Co dynamics: from ${\cal X}$ to ${\cal Y}_{3}$, all feature-categories resulted from applying the Splitting-Stopping-rule(SSR) algorithm are defined and reported via HC-trees of all orders in Fig~\ref{SDAtreeseveTovi} and via 1D and 2D scatter plots in Fig~\ref{SDAplotseveTovi}. While Iris' binary memberships of all confirmed major feature-categories are collectively reported through a heatmap in Fig~\ref{seveTovi}.

Upon the pairwise comparison of $setosa-vs-vesicolor$, the heatmap of Fig~\ref{seveTovi} shows a HC-tree with three major branches: $L1$, $R1L2$ and $R1R2$. The two major branches: $L1$ exclusively for $setosa$ and $R1L2$ containing all 50 $versicolor$, are balanced in the sense of being characterized by two complementary horizontal chains of block of complete 1s and 0s. Both blocks contain all 12 computed and confirmed major feature-categories of all orders involving either Petal.Width or Petal.Length. Both Septal.Length and Septal.Width are missing. The 44 ``unknown'' $virginica$ Irises' in branch $R1R2$ are apparent outliers, while outlier statuses of the 6 ``unknown'' $virginica$ Irises' in branch $R1L2$ are to some degree visible. Such pieces of information have multiple implications in classifications between $versicolor$ and $virginica$. This is the merit of Class-Informatics.

\subsection{Class-informatics via ${\cal Y}_{2}$ }
\begin{figure}[h!]
 \centering
\includegraphics[width=0.8\textwidth]{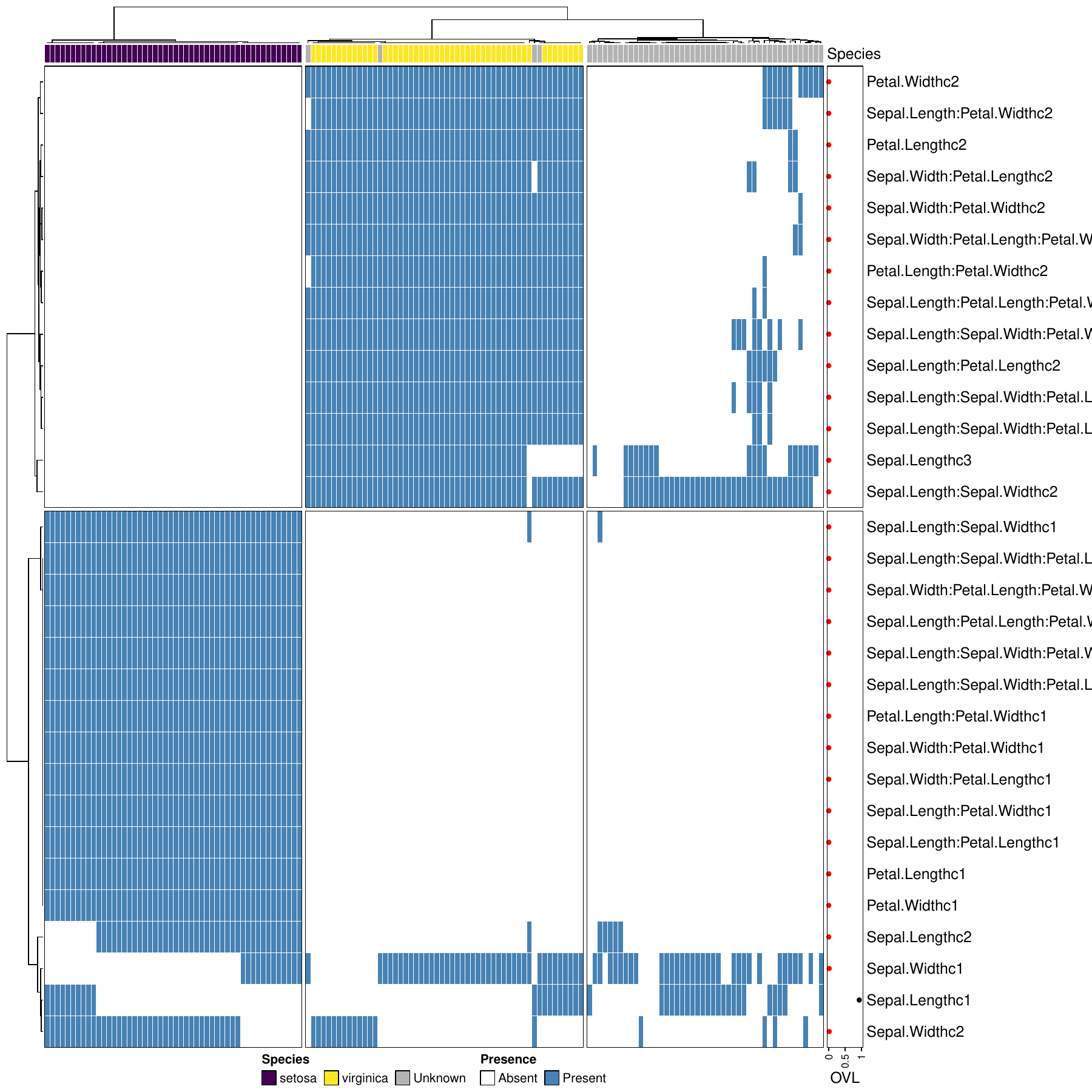}
 \caption{heatmap of classification on $\{setosa, virginica\}$ all major feature-categories from order one to four coupled with $versicolor$'s information outlier status. $versicolor$ Irises are marked as ``unknown''.}
\label{seviTove}
 \end{figure}

\begin{figure}[h!]
 \centering
\includegraphics[width=1.0\textwidth]{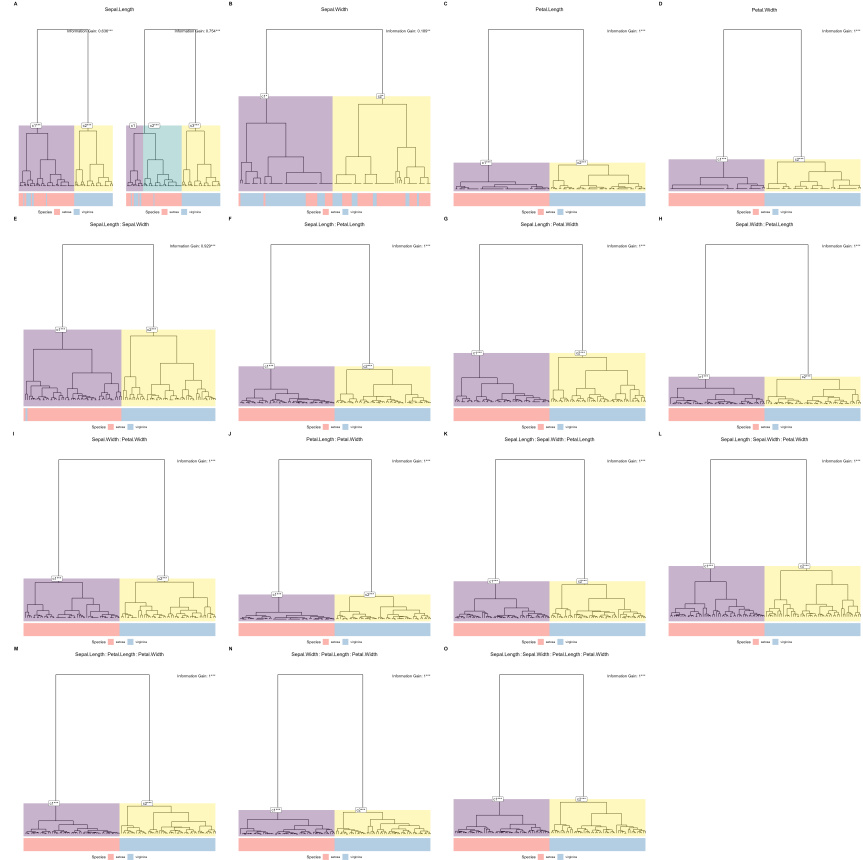}
 \caption{SDA computational results on $\{setosa, virginica\}$ with all major feature-categories from order one to four. Each major feature-category's category-codes $ck$ with $k=1,2...$ via HC-trees are explicitly listed. All columns of Heatmap in Fig~\ref{seviTove} are defined.}
\label{SDAtreeseviTove}
 \end{figure}

\begin{figure}[h!]
 \centering
\includegraphics[width=1.0\textwidth]{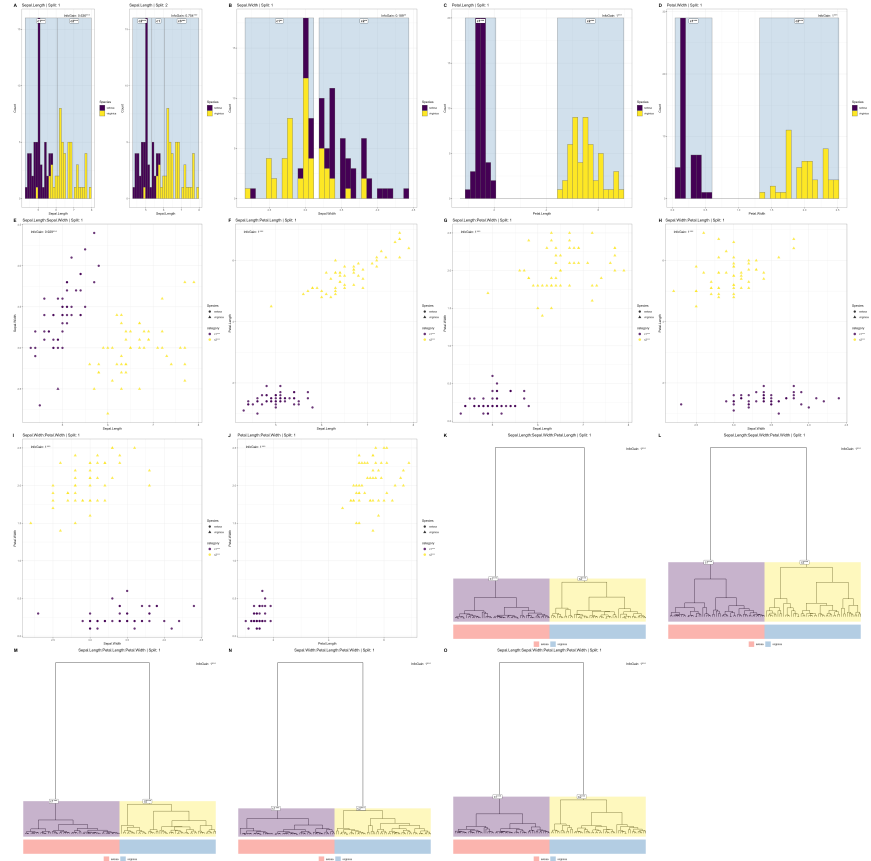}
 \caption{SDA computational results on $\{setosa, virginica\}$ with all major feature-categories from order one to four. Each major feature-category's precise geometric definitions and biological meanings of category-codes $ck$ with $k=1,2...$ are explicitly listed. All columns of Heatmap in Fig~\ref{seviTove} are explicitly explained.}
\label{SDAplotseviTove}
 \end{figure}

Finally, we start subsystem-explorations via the Re-Co dynamics: from ${\cal X}$ to ${\cal Y}_{2}$, with ${\cal Y}_{2}=\{Se, Vi\}$. Within the Re-Co dynamics: from ${\cal X}$ to ${\cal Y}_{2}$, all feature-categories resulted from applying the Splitting-Stopping-rule(SSR) algorithm are defined and reported via HC-trees of all orders in Fig~\ref{SDAtreeseviTove} and via 1D and 2D scatter plots in Fig~\ref{SDAplotseviTove}. While Iris' binary memberships of all confirmed major feature-categories are collectively reported through a heatmap in Fig~\ref{seviTove}.

The heatmap for pairwise comparison ${cal Y}_{2}$ for $setosa-vs-virginica$, as shown in Fig~\ref{seviTove}, delivers very much corresponding block-patterns for characterizing the three class-informatics by switching the roles of $virginia$ and $vesicolor$ regarding that in Fig~\ref{seveTovi}. In particular, the two horizontal block-chains framed by two branches $\{T1, B1T2\}$ on the row-axis and three branches $\{L1, R1L2, R1R2\}$ on the column-axis are rather correspondingly similar. Especially, when focuses are placed on the 12 computed and confirmed major feature-categories of all orders involving either Petal.Width or Petal.Length, the block structures are almost identical. With such focuses, these two block-chains of Fig~\ref{seviTove} clearly demonstrates the information outlier status for Irises of $versicolor$ with respect to the pairwise comparison of $setosa-vs-virginica$, except one. As such Irises of $versicolor$-informatics are precisely described within the $setosa-vs-virginica$ comparison. Nonetheless, this one $versicolor$ is separated from all $virginia$ in Fig~\ref{seveTovi}. Therefore, they would not be mistaken as $virginia$ when these two figures are put together.

After all block-based large scale patterns narrated and explained through heatmaps resulted from the ${\cal Y}=\{Se, Ve, Vi\}$ and three pairwise comparisons: ${\cal Y}_1=\{Ve, Vi\}$, ${\cal Y}_2=\{Se, Vi\}$ and ${\cal Y}_3=\{Se, Ve\}$, it is essential to turn to fine scale information contained in the three class-informatics. Since each heatmap defines a topological space for iris-subjects involving in its row-axis. Via each individual character-landscape, we see which Irises are neighbors of which Irises. In contrast, via binary memberships of each major feature-category, we see which major feature-categories are neighbors of which feature-categories. Even though neighbors might be of different orders, their mechanistic dependency is the core materials for motivating better understanding, brand-new knowledge and possibly unexpected intelligence.

Such fine scale topological information of all Iris and all major feature-categories provide the basis for inferential decision-making regarding any unknown subject. That is, all these topological spaces are to be explored by any unknown Iris in individual and multiscale neighborhood bases. Such explorations going through the three species-specific class-informatics would collectively provide very precise insights to facilitate a proper inferential decision coupled with very good explanations. This inferential function is demonstrated as a by-product of systemic Class-informatics. Once again, the most essential goal of any scientific study on any ``so-called classification data set'' is indeed to build its Class-informatics for its taxonomy. Here, this is how we have achieved this goal on Iris data via SDA.

\section{Conclusions: SDA based explanatory Class-informatics}
Across the full Re-Co dynamics from ${\cal X}$ to ${\cal Y}$ and the spectrum of subsystem Re-Co dynamics from ${\cal X}$ to ${\cal Y}_{sub}$, the four sets of heatmaps reported in the two previous section constitute the SDA based explanatory Class-informatics of Iris data. Such Class-informatics  carries a rather wide spectrum of aspects of characterizations for all 150 Iris' individual character-landscapes and topological neighborhood systems. And it also simultaneously carries a rather wide spectrum of potential mechanistic dependence among major and interacting effects of all orders pertaining to all four covariate features. As such this Class-informatics based taxonomy of Iris is gravitated toward the fixed collection of all possible 12 features and feature-sets involving with either Petal.Length or Petal.Width. In contrast, the Sepal.Width by itself is acting like a noise, unless when it is coupled with other three features.

Specifically, each study subject Iris is characterized via four versions of Re-Co dynamics specific individual character-landscapes together with a collective of individual outlier status across confirmed major features-categories. As such each individual study subject can be very precisely described with its idiosyncratic characteristics in detail. With respect to a heatmap specific topology, each subject will be prescribed with pattern information regarding which subjects are its neighbors, which subjects are contrastingly distinct in characters of fine and large scales. All such descriptions are scientific explanations. It is noted that such explanations are subject to choices of signal strength and finite sample precisions supported by data. This key part of Class-informatics is what a knowledgeable taxonomist is supposed to do.

Further, Class-informatics would and should facilitate explanations on heatmap specific block-based mechanistic dependence among major (order-1) and interacting effects (order-2 and higher) of all involved covariate features. Since a block in a heatmap is framed by a cluster of study subjects and a cluster of major feature-categories of various orders. By sharing a common set of study subjects, this cluster of major feature-categories indeed pinpoint scientific patterns aiming for biological mechanisms that could be highly important and potential for expanding current stage of knowledge regarding the complex system under study. Across all heatmaps, the collective of such patterns of mechanistic dependence might be beyond any knowledgeable taxonomist's subject matter intelligence.

The implications of Class-informatics on predictive inferential decision-making are transparently clear. Individual outlier detection tasks are needed across the full Re-Co dynamics from ${\cal X}$ to ${\cal Y}$ and the spectrum of subsystem Re-Co dynamics from ${\cal X}$ to ${\cal Y}_{sub}$.
Since, when observed data points of study subjects are marked with annotated response-ID in ${\cal Y}^{(C)}_{sub}$, the rest of data points are marked with response-IDs of ${\cal Y}_{sub}$. The role of the targeted yet-to-be predicted subject $x^@$ would need to be consistent with some observed data points' nearest neighborhood to make its prediction scientifically valid. And such a prediction is explained as such. That is, its geometric or information outlier statuses are critical Classification information. In other words, the outlier status is a critical source of Classification information. This perspective of Classification is rather essential in taxonomy as a key part of sciences. This is another aspect of merits of Class-informatics.

One rather unusual merit of Class-informatics is: Annotation-correction. An observed data point is annotated with a response-ID. When it is within a subsystem ${\cal Y}_{sub}$, it behaves like an information outlier. While it is not participating within a subsystem ${\cal Y}_{sub}$, it behaves as neither a geometric nor an information outlier. Such a data point attached with such ``erroneous behaviors'' would surely strike a loud warning of potential ``mislabeling'' or ``wrong-annotation''. Such kinds of errors are not uncommon especially when the IDs of ${\cal Y}$ are defined in a data-driven fashion, such as employing HC-algorithm among many others. Such errors certainly are unavoidable due to unperfect expert knowledge or instrumental measurements. This is one of key merits of Class-informatics. So, this is what a knowledgeable taxonomist should be capable of.

Finally, we mention one critical merit of Class-informatics via SDA computing perspective: Checking the equality of information contents between training and testing data-split on the original data set. This is the most fundamental assumption in Machine Learning and Deep Learning. The quest of checking this assumption has not been done in literature. Our resolution to this quest is: SDA computing. Two key SDA computing steps are outlined here for testing this equality assumption. First, all testing data points can not be either information or geometric outliers under all Re-Co dynamics from ${\cal X}$ to ${\cal Y}$ and the spectrum of subsystem Re-Co dynamics from ${\cal X}$ to ${\cal Y}_{sub}$. Secondly, all testing data points taking the role of non-participants must coherently and correctly emerge with patterns of all training data points' individual character-landscapes of all scales pertaining to topological systems of any Re-Co dynamics. Any confirmed outlier status would be a piece of negative information against the equality assumption. This resolution to testing this assumption is the critical merit of Class-informatics in Machine Learning and Deep Learning literature.

\end{document}